%% file: PBH.tex
\documentclass[a4paper,11pt]{article}
\pdfoutput=1 
\usepackage{jcappub}
\usepackage{amsmath}
\usepackage{bm}
\usepackage{graphicx}
\usepackage{enumitem}
\usepackage{geometry}
\usepackage{xspace}
\usepackage{pdflscape}

\makeatletter
\gdef\@fpheader{}
\g@addto@macro\bfseries{\boldmath}
\makeatother

\input{newcommands}

\subheader{}

\title{Quantum diffusion during inflation and primordial black holes}

\author[a]{Chris Pattison,}
\author[b,a]{Vincent Vennin,}
\author[a,c]{Hooshyar Assadullahi,}
\author[a]{and David Wands}

\affiliation[a]{Institute of Cosmology \& Gravitation, University of Portsmouth, Dennis Sciama Building, Burnaby Road, Portsmouth, PO1 3FX, United Kingdom}

\affiliation[b]{Laboratoire Astroparticule et Cosmologie, Universit\'e Denis Diderot Paris 7, 75013 Paris,
France}

\affiliation[c]{School of Earth and Environmental Sciences, University of Portsmouth, Burnaby Building, Burnaby Road, Portsmouth, PO1 3QL, United Kingdom}

\emailAdd{hooshyar.assadullahi@port.ac.uk}
\emailAdd{christopher.pattison@port.ac.uk}
\emailAdd{vincent.vennin@port.ac.uk}
\emailAdd{david.wands@port.ac.uk}

\date{today}

\begin{document}
\sloppy

\abstract{We calculate the full probability density function (PDF) of inflationary curvature perturbations, even in the presence of large quantum backreaction. Making use of the stochastic-$\delta N$ formalism, two complementary methods are developed, one based on solving an ordinary differential equation for the characteristic function of the PDF, and the other based on solving a heat equation for the PDF directly. In the classical limit where quantum diffusion is small, we develop an expansion scheme that not only recovers the standard Gaussian PDF at leading order, but also allows us to calculate the first non-Gaussian corrections to the usual result. In the opposite limit where quantum diffusion is large, we find that the PDF is given by an elliptic theta function, which is fully characterised by the ratio between the squared width and height (in Planck mass units) of the region where stochastic effects dominate. We then apply these results to the calculation of the mass fraction of primordial black holes from inflation, and show that no more than $\sim 1$ $e$-fold can be spent in regions of the potential dominated by quantum diffusion. We explain how this requirement constrains inflationary potentials with two examples.}


\arxivnumber{1707.00537}

\maketitle

\section{Introduction and motivations}
\label{sec:Intro}
Cosmological inflation \cite{Starobinsky:1980te, Sato:1980yn, Guth:1980zm, Linde:1981mu, Albrecht:1982wi, Linde:1983gd} is a period of accelerated expansion that occurred at very high energy in the early Universe. During this epoch, vacuum quantum fluctuations were amplified to become large-scale cosmological perturbations that seeded the cosmic microwave background (CMB) anisotropies and the large-scale structure of our Universe \cite{Mukhanov:1981xt, Mukhanov:1982nu, Starobinsky:1982ee, Guth:1982ec, Hawking:1982cz, Bardeen:1983qw}.

In the range of scales accessible to CMB experiments~\cite{Ade:2015xua, Ade:2015lrj}, these perturbations are constrained to be small, at the level $\zeta\simeq 10^{-5}$ until they re-enter the Hubble radius during the radiation era, where $\zeta$ is the scalar curvature perturbation. At smaller scales however, they may be sufficiently large so that when they re-enter the Hubble radius, they overcome the pressure forces and collapse to form primordial black holes (PBHs)~\cite{Hawking:1971ei, Carr:1974nx, Carr:1975qj}. In practice, PBHs form when the mean curvature perturbation in a given Hubble patch exceeds a threshold denoted $\zeta_\uc\simeq 1$~\cite{Zaballa:2006kh, Harada:2013epa} (see \Ref{Young:2014ana} for an alternative criterion based on the density contrast rather than the curvature perturbation). 

The abundance of PBHs is usually stated in terms of the mass fraction of the Universe contained within PBHs at the time of formation, $\beta$. If the coarse-grained curvature perturbation $\zeta_{\mathrm{cg}}$ follows the probability distribution function (PDF) $P(\zeta_{\mathrm{cg}})$, $\beta$ is given by~\cite{1975ApJ...201....1C}
\bea
\label{eq:beta:def}
\beta\left(M\right) = 2\int_{\zeta_\uc}^\infty P\left(\zeta_{\mathrm{cg}}\right) \dd \zeta_{\mathrm{cg}}\, .
\eea
Here, $\zeta_{\mathrm{cg}}$ is obtained from keeping the wavelengths smaller than the Hubble radius at the time of formation, 
\bea
\label{eq:zetacg:def}
\zeta_{\mathrm{cg}}(\bm{x}) = \left(2\pi\right)^{-3/2}\int_{k>aH_{\mathrm{form}}}\dd {\bm{k}}\zeta_{\bm{k}} e^{i\bm{k}\cdot\bm{x}}\, ,
\eea
where $a$ is the scale factor, $H\equiv \dot{a}/a$ is the Hubble scale, and a dot denotes differentiation with respect to cosmic time. In \Eq{eq:beta:def}, $M$ is the mass contained in a Hubble patch at the time of formation~\cite{Choptuik:1992jv, Niemeyer:1997mt, Kuhnel:2015vtw}, $M=3\Mp^2/H_{\mathrm{form}}$, where $\Mp$ is the reduced Planck mass.

Observational constraints on $\beta$ depend on the masses PBHs have when they form. For masses between $10^9\mathrm{g}$ and $10^{16} \mathrm{g}$, the constraints mostly come from the effects of PBH evaporation on big bang nucleosynthesis and the extragalactic photon background, and typically range from $\beta<10^{-24}$ to $\beta<10^{-17}$. Heavier PBHs, with mass between $10^{16} \mathrm{g}$ and $10^{50} \mathrm{g}$, have not evaporated yet and can only be constrained by their gravitational and astrophysical effects, at the level $\beta<10^{-11}$ to $\beta<10^{-5}$ (see \Refs{Carr:2009jm, Carr:2017jsz} for summaries of constraints).

Compared to the CMB anisotropies that allow one to measure $\zeta$ accurately in the largest $\sim 7$ \efolds~of scales in the observable Universe, PBHs only provide upper bounds on $\beta(M)$, and hence on $\zeta$. However, these constraints span a much larger range of scales and therefore yield valuable additional information. This is why PBHs can be used to constrain the shape of the inflationary potential beyond the $\sim 7$ \efolds~that are accessible through the CMB. 

In practice, one usually assumes $P(\zeta_{\mathrm{cg}})$ to be a Gaussian PDF with standard deviation given by the integrated power spectrum $\left\langle \zeta_{\mathrm{cg}}^2 \right\rangle = \int_{k}^{k_\uend} \calP_\zeta(\tilde{k})\dd \ln \tilde{k}$, where $k$ is related to the time of formation through $k=aH_{\mathrm{form}}$, and where $k_\uend$ corresponds to the wavenumber that exits the Hubble radius at the end of inflation. Combined with \Eq{eq:beta:def}, this gives rise to
\bea
\label{eq:beta:erfc}
\beta\left(M\right) = \erfc\left[\frac{\zeta_\uc}{\sqrt{2  \int_{k}^{k_\uend} \calP_\zeta(\tilde{k})\dd \ln \tilde{k}}}\right]\, ,
\eea
where $\erfc$ is the complementary error function, $M$ is the mass contained in the Hubble volume, and $2\pi/k$ is the comoving Hubble length when the black holes form. In the limit $\beta \ll 1$, this leads to $\int_{k}^{k_\uend} \calP_\zeta(\tilde{k})\dd \ln \tilde{k} \simeq \zeta_\uc^2/(-2\ln\beta)$. Assuming the power spectrum to be scale invariant, one has $\int_{k}^{k_\uend} \calP_\zeta(\tilde{k})\dd \ln \tilde{k} \simeq \calP_\zeta \ln(k_\uend/k)  \simeq \calP_\zeta \Delta N$, where $\Delta N = \Delta \ln a$ is the number of \efolds~elapsed between the Hubble radius exit times of $k$ and $k_\uend$ during inflation. This leads to
\bea
\label{eq:Pzetaconstraint:standard}
\calP_\zeta \Delta N \simeq - \frac{\zeta_\uc^2}{2\ln\beta}\, .
\eea
For instance, with $\zeta_\uc=1$, the bound $\beta<10^{-22}$ leads to the requirement that $\calP_\zeta \Delta N< 10^{-2}$. This can be translated into constraints on the inflationary potential $V=24\pi^2\Mp^4 v$ and its derivative $V^\prime$ with respect to the inflaton field $\phi$ using the slow-roll formulae~\cite{Mukhanov:1985rz, Mukhanov:1988jd} 
\bea
\label{eq:classicalPS}
\calP_\zeta = \frac{2v^3}{\Mp^2 {v^\prime}^2}\, ,\quad 
\Delta N=\int_{\phi_\uend}^\phi\frac{v}{\Mp^2v'}   \dd \tilde{\phi} \, .
\eea

The crucial remark that motivates the present work is that the assumptions on which the above considerations rely, namely the use of a Gaussian PDF for $P(\zeta_{\mathrm{cg}})$ together with the classical slow-roll formula for the curvature power spectrum $\calP_\zeta$ and number of \efolds~$\Delta N$, are valid only in the regime where quantum diffusion provides a subdominant correction to the classical field dynamics during inflation. However, as we shall now see, producing curvature fluctuations of order $\zeta\sim \zeta_\uc\sim 1$ or higher precisely corresponds to the regime where quantum diffusion dominates the field dynamics over a typical time scale of one \efold. The validity of the standard approach that is summarised above is therefore questionable and this is why in this paper, we present a generic calculation of the PBH abundance from inflation that fully incorporates quantum backreaction effects, and we update \Eq{eq:Pzetaconstraint:standard} to take into account the full quantum dynamics of the inflaton field.

In practice, we make use of the stochastic inflation formalism~\cite{Starobinsky:1986fx, Nambu:1987ef, Nambu:1988je, PhysRevD.39.2245, Nakao:1988yi, Nambu:1989uf, Mollerach:1990zf, Linde:1993xx, Starobinsky:1994bd}, which is an effective theory for the long-wavelength parts of the quantum fields during inflation. When light fields are coarse grained at a fixed, non-expanding, physical scale that is larger than the Hubble radius during the whole period of inflation, one can show  that their dynamics indeed become classical and stochastic. In the slow-roll approximation, the inflaton field $\phi$ follows a Langevin equation of the form
\bea
\label{eq:intro:Langevin}
\frac{\dd \phi}{\dd N} = -\frac{V'}{3H^2} + \frac{H}{2\pi}\xi\left( N \right) \, .
\eea
The right-hand side of this equation has two terms, the first of which involves $V'$ and is a classical drift term, and the second term involves $\xi$ which is a Gaussian white noise such that $\left\langle \xi \left( N \right) \right\rangle = 0$ and $\left\langle \xi \left( N \right) \xi \left( N' \right)\right\rangle = \delta\left( N - N' \right)$, and which makes the dynamics stochastic.

Over the time scale of one \efold, the ratio between the mean quantum kick, $H/(2\pi)$, and the classical drift, $V^\prime/(3H^2)$, is of order $\sqrt{\calP_\zeta}$, provided $\calP_\zeta$ follows the classical formula~(\ref{eq:classicalPS}) and where one has made use of the Friedmann slow-roll equation $H^2\simeq 8\pi^2\Mp^2 v$. Therefore, if PBHs form when this ratio is of order one or higher, this is precisely when one expects quantum modifications to the standard result to become important. 

The effects of quantum diffusion on  PBHs formation can thus be dramatic and are the subject of this paper, which is organised as follows. In \Sec{sec:PDF}, we explain how the full PDF of curvature perturbations can be calculated in stochastic inflation. Using the stochastic-$\delta N$ formalism (see \Sec{sec:stochasticdeltaN}), we first derive a set of ordinary differential equations for the moments of this PDF (see \Sec{sec:FPTmoments}), from which two methods of construction of the distribution are proposed, one based on its characteristic function (see \Sec{sec:CharacteristicFunction}) and one based on a heat equation (see \Sec{sec:heatequation}). In \Sec{sec:ClassicalLimit}, we derive the classical limit of our formulation, where quantum diffusion is a subdominant correction to the classical field dynamics. At leading order, the standard result is recovered, and higher-order corrections allow us to calculate the first non-Gaussian modifications to the PDF of curvature perturbations and to the mass fraction $\beta$. In \Sec{sec:StochasticLimit}, we expand our calculation in the opposite limit, where the potential is exactly flat and stochastic effects dominate. In this case, the PDF of curvature perturbations is found to be highly non-Gaussian and is given by an elliptic theta function. In \Sec{sec:PBH}, we explain how these two limits enable one to treat more generic inflationary potentials and give a simple calculational programme that updates \Eq{eq:Pzetaconstraint:standard} and allows one to translate PBH observational constraints into constraints on the potential. We then illustrate this programme with two examples. We finally summarise our main results and present our conclusions in \Sec{sec:Conclusion}.
\section{Probability distribution of curvature perturbations}
\label{sec:PDF}
The calculation of the PBH mass fraction relies on the PDF of the coarse-grained curvature perturbations through \Eq{eq:beta:def}. Let us explain how this distribution can be calculated in the stochastic-$\delta N$ formalism.
\subsection{The stochastic-$\delta N$ formalism}
\label{sec:stochasticdeltaN}
\subsubsection{The $\delta N$ formalism}
The starting point of the stochastic-$\delta N$ formalism is the standard, classical $\delta N$ formalism~\cite{Starobinsky:1982ee, Starobinsky:1986fxa, Sasaki:1995aw, Sasaki:1998ug, Lyth:2004gb, Lyth:2005fi}, which provides a succinct way of relating the fluctuations in the number of \efolds~of expansion during inflation for a family of homogeneous universes with the statistical properties of curvature perturbations. Starting from the unperturbed flat Friedmann-Lema\^{i}tre-Robertson-Walker metric
\bea
\label{eq:metric:FLRW}
\dd s^2 = -\dd t^2 + a^2(t)\delta_{ij}\dd x^{i}\dd x^{j} \, ,
\eea
deviations from isotropy and homogeneity can be added at the perturbative level and contain scalar, vector and tensor degrees of freedom. Gauge redundancies associated with diffeomorphism invariance allow one to choose a specific gauge in which fixed time slices have uniform energy density and fixed spatial worldlines are comoving (in the super-Hubble regime this gauge coincides with the synchronous gauge supplemented by some additional conditions that fix it uniquely). Including spatial perturbations only, one obtains~\cite{Starobinsky:1982ee, Creminelli:2004yq,Salopek:1990jq}
\bea
\dd s^2 = -\dd t^2 + a^2(t)\ee^{2\zeta(t, \bm{x})}\delta_{ij}\dd x^{i}\dd x^{j} \, ,
\eea
where $\zeta$ is the adiabatic curvature perturbation mentioned in \Sec{sec:Intro}. One can then introduce a local scale factor
\bea
\label{eq:alocal:def}
\tilde{a}(t, \bm{x}) = a(t)\ee^{\zeta(t, \bm{x})} \, ,
\eea
which allows us to express the amount of expansion from an initial flat space-time slice at time $t_\uin$ to a final space-time slice of uniform energy density as
\bea
N(t, \bm{x}) = \ln{\left[ \frac{\tilde{a}(t, \bm{x})}{a(t_{\mathrm{in}})} \right]} \, .
\eea
This is related to the curvature perturbation $\zeta$ via \Eq{eq:alocal:def}, which gives rise to
\bea
\label{eq:zeta:deltaN}
\zeta(t, \bm{x}) = N(t, \bm{x}) - \bar{N}(t) \equiv \delta N \, ,
\eea
where $\bar{N}(t) \equiv \ln{\left[ {a(t)}/{a(t_{\mathrm{in}})} \right]}$ is the unperturbed expansion. 
This expression forms the basis of the $\delta N$ formalism, which follows by making the further simplifying assumption that on super-Hubble scales, each spatial point of the universe evolves independently and this evolution is well approximated by the evolution of an unperturbed universe.  This assumption is known as the ``quasi-isotropic''~\cite{Lifshitz:1960, Starobinsky:1982mr, Comer:1994np, Khalatnikov:2002kn} or ``separate universe'' approach~\cite{Wands:2000dp, Lyth:2003im, Lyth:2004gb}, and allows us to neglect spatial gradients on super-Hubble scales. As a consequence, $N(t, \bm{x})$ is the amount of expansion in unperturbed, homogeneous universes, and $\zeta$ can be calculated from the knowledge of the evolution of a family of such universes.
\subsubsection{The stochastic-$\delta N$ formalism}
\begin{figure}[t]
\begin{center}
\includegraphics[width=0.6\textwidth]{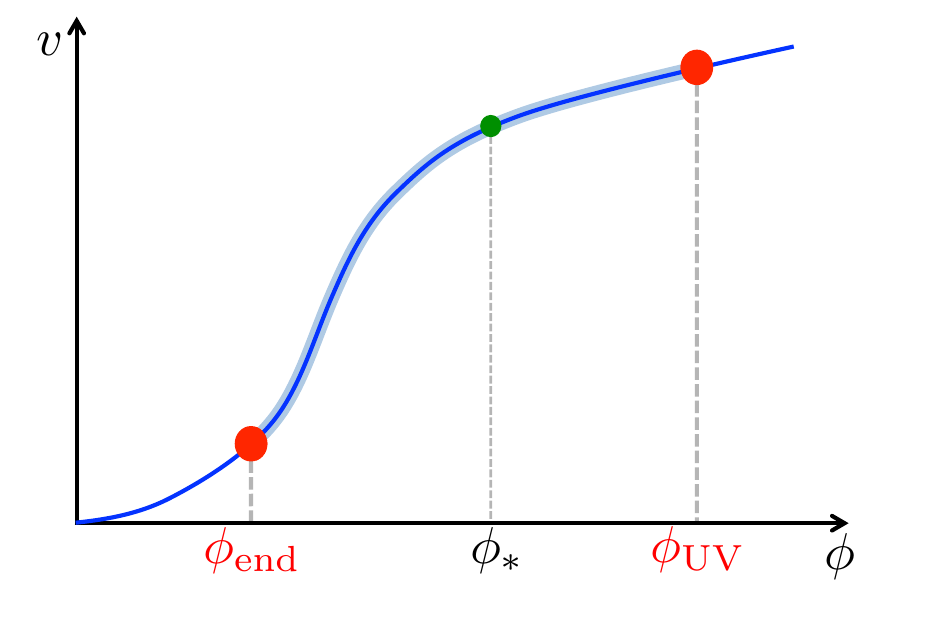}
\caption{Sketch of the single-field stochastic dynamics solved in this work. Starting from $\phi_*$ at initial time, the inflaton field $\phi$ evolved along the potential $v(\phi)$ under the Langevin equation~(\ref{eq:intro:Langevin}), until $\phi$ reaches $\phi_\uend$ where inflation ends. A reflective wall is added at $\phiuv$ to prevent the field from exploring arbitrarily large values. The number of \efolds~realised along a family of realisations of the Langevin equation is calculated, and gives rise to the probability distribution of curvature perturbations using the $\delta N$ formalism.}  
\label{fig:sketch}
\end{center}
\end{figure}
The $\delta N$ formalism relies on the calculation of the amount of expansion realised amongst a family of homogeneous universes. When stochastic inflation is employed to describe such a family of universes and to calculate the amount of expansion realised in them, this gives rise to the stochastic-$\delta N$ formalism~\cite{Enqvist:2008kt, Fujita:2013cna, Fujita:2014tja, Vennin:2015hra, Kawasaki:2015ppx, Assadullahi:2016gkk, Vennin:2016wnk}.

This approach is sketched in \Fig{fig:sketch}. Starting from $\phi=\phi_*$ at an initial time, the inflaton field evolves along the potential $v(\phi)$ under the Langevin equation~(\ref{eq:intro:Langevin}), where hereafter we use the rescaled dimensionless potential
\bea
v\equiv \frac{V}{24\pi^2\Mp^4}\, ,
\eea
until it reaches $\phi_\uend$ where inflation ends. A reflective wall is added at $\phiuv$ to prevent the field from exploring arbitrarily large values, which can be necessary to renormalise infinities appearing in the theory~\cite{Vennin:2016wnk} (whose results are still independent of $\phiuv$ and of the exact nature of the wall, reflective or absorbing, provided $\phiuv$ lies in some range). The amount of expansion realised along a given trajectory is called $\mathcal{N}$, which is a stochastic variable. Thanks to the $\delta N$ formalism, the fluctuation in this number of \efolds, $\mathcal{N}-\langle \mathcal{N} \rangle$, is nothing but the coarse-grained curvature perturbation $\zeta_{\mathrm{cg}}$ defined in \Eq{eq:zetacg:def},
\bea
\delta N_{\mathrm{cg}}\left(\bm{x}\right) = 
\mathcal{N}\left(\bm{x}\right)-\left\langle \mathcal{N} \right\rangle =
\zeta_{\mathrm{cg}}\left(\bm{x}\right) = 
\frac{1}{\left(2\pi\right)^{3/2}}
\int_{k_*}^{k_\uend} \dd \bm{k} \zeta_{\bm{k}} e^{i \bm{k}\cdot \bm{x}}\, ,
\eea
where $k_*$ and $k_\uend$ are the wavenumbers that cross the Hubble radius at initial and final times when $\phi=\phi_*$ and $\phi=\phi_\uend$ respectively. Since, as explained in \Sec{sec:Intro}, the calculation of the PBH mass function relies on the PDF of coarse-grained curvature perturbations, the next step is to calculate the PDF of $\delta N_\mathrm{cg}$.

Before doing so, let us note that quantities related to $\zeta$, and not $\zeta_{\mathrm{cg}}$, can also be calculated in the stochastic-$\delta N$ formalism. For the power spectrum $\calP_\zeta$ for instance, since the coarse-grained $\delta N_{\mathrm{cg}}$ receives an integrated contribution of all modes exiting the Hubble radius during inflation, $\langle \delta N_\mathrm{cg}^2 \rangle = \int_{k_*}^{k_\uend} \calP_\zeta \dd k/k$, one has~\cite{Fujita:2013cna, Vennin:2015hra}
\bea
\label{eq:Pzeta:stochaDeltaN}
\calP_\zeta = \frac{\dd \left\langle \delta N_{\mathrm{cg}}^2 \right\rangle}{\dd \left\langle \mathcal{N}\right\rangle}\, ,
\eea
where we have used the relation $\langle \mathcal{N}\rangle = \ln(a_\uend/a_*) = \ln(k_\uend/k)$, where the last equality is valid at leading order in slow roll only. In the same manner, the local bispectrum can be written as $\mathcal{B}_\zeta \propto \dd^2 \langle \delta N_\mathrm{cg}^3\rangle/\dd \langle \mathcal{N} \rangle^2$, from which the effective $\fnl^\mathrm{local}$ parameter, measuring the ratio between the bispectrum and the power spectrum squared, is given by
\bea
\label{eq:fnl:stochaDeltaN}
\fnl^\mathrm{local} = \frac{5}{72}\frac{\left\langle \delta N_\mathrm{cg}^3\right\rangle}{\dd \left\langle \mathcal{N} \right\rangle^2}\left(\frac{\dd \left\langle \delta N_{\mathrm{cg}}^2 \right\rangle}{\dd \left\langle \mathcal{N}\right\rangle}\right)^{-2}\, .
\eea
\subsection{Statistical moments of first passage times}
\label{sec:FPTmoments}
In order to calculate the PDF of the realised number of \efolds~$\N$, and hence of $\delta N_\mathrm{cg}$ (\ie of $\zeta_\mathrm{cg}$), a first step consists in calculating its statistical moments
\bea
\label{eq:moments}
f_n(\phi) = \left\langle \N^n(\phi) \right\rangle\, ,
\eea
where the dependence on the field value $\phi$ (denoted $\phi_*$ in the discussion around \Fig{fig:sketch}) at which trajectories are initiated is made explicit. This can be done using the techniques of ``first passage time analysis''~\cite{Bachelier:1900, Gihman:1972}, applied to stochastic inflation in \Refs{Starobinsky:1986fx, Vennin:2015hra}, which allow one to derive a hierarchy of ordinary differential equations
\bea
\label{eq:ODE:fn}
f_n^{\prime\prime} - \frac{v^\prime}{v^2} f_n^\prime = -\frac{n}{v\Mp^2} f_{n-1}\, .
\eea
The hierarchy is initiated at $f_0 = 1$, and for $n \geq 1$ it has to be solved with two boundary conditions, one related to the fact that all trajectories initiated at $\phi_\uend$ realise a vanishing number of \efolds, and the other one implementing the presence of a reflective wall at $\phiuv$, namely
\bea
\label{eq:BC}
f_n(\phiend) = 0\, ,\quad f_n^\prime(\phiuv) = 0\, .
\eea
The formal solution to this problem can be written as
\bea
\label{eq:fn:generalsolution}
f_n(\phi) = n \int_{\phiend}^\phi \frac{\dd x}{\Mp} \int_x^{\phiuv} \frac{\dd y}{\Mp} \ee^{\frac{1}{v(y)}-\frac{1}{v(x)}} \frac{f_{n-1}(y)}{v(y)}\, ,
\eea
which allows one to calculate the moments iteratively. In practice, this relies on performing integrals of increasing dimension, which quickly becomes numerically heavy but provides a convenient way to study the first few moments required to calculate the power spectrum given by \Eq{eq:Pzeta:stochaDeltaN} or the $\fnl^\mathrm{local}$ parameter given by \Eq{eq:fnl:stochaDeltaN}, see \Refs{Vennin:2015hra, Vennin:2016wnk}.
\subsection{The characteristic function approach}
\label{sec:CharacteristicFunction}
In order to relate the PDF of $\N$ to its statistical moments, let us introduce its characteristic function
\bea
\label{eq:characteristicFunction:def}
\chi_\N(t,\phi) \equiv \left\langle \ee^{it \N(\phi)} \right\rangle \, ,
\eea 
which depends on $\phi$ and a dummy parameter $t$. By Taylor expanding $\chi_\N(t,\phi)$ around $t=0$, one has $\chi_\N(t,\phi) = \sum_{n=0}^\infty (it)^n f_n(\phi)/n!$. If one applies the differential operator appearing in the left-hand side of \Eq{eq:ODE:fn} to this expansion, and uses \Eq{eq:ODE:fn}  to replace each term by its right-hand side, one obtains
\bea
\label{eq:ODE:chi}
\left(\frac{\partial^2}{\partial\phi^2} - \frac{v^\prime}{v^2} \frac{\partial}{\partial\phi} + \frac{it}{v\Mp^2}\right)\chi_\N(t,\phi) = 0 \, .
\eea
At fixed $t$, this is an ordinary differential equation in $\phi$, so instead of the hierarchy of coupled differential equations~(\ref{eq:ODE:fn}) one now has a set of uncoupled differential equations to solve, which improves the tractability of the problem. The boundary conditions~(\ref{eq:BC}), together with the fact that $f_0=1$, translate into
\bea
\label{eq:boundary:chi}
\chi_\N\left(t,\phiend\right) = 1\, ,\quad
\frac{\partial\chi_\N}{\partial\phi} \left(t, \phiuv\right) = 0 \, . 
\eea
Let us note that the characteristic function of the fluctuation in the number of \efolds, $\zeta_\mathrm{cg} = \delta N_\mathrm{cg} = \N - \langle \N \rangle = \N -f_1$, can be found by plugging this expression into \Eq{eq:characteristicFunction:def}, which gives rise to 
\bea
\label{eq:chideltaN:chiN}
\chi_{\zeta_\mathrm{cg}}\left(t,\phi\right) = e^{-i f_1(\phi) t} \chi_\N(t,\phi)\, .
\eea
Finally, from  \Eq{eq:characteristicFunction:def}, the characteristic function $\chi_\N$ can be rewritten as
\bea
\label{eq:chi:P}
\chi_{\N}(t,\phi)=\int^{\infty}_{-\infty} \ee^{it\N} P\left(\N, \phi\right) \dd \N\, ,
\eea
that is to say, the characteristic function is the Fourier transform of the PDF of curvature perturbations. Therefore, the PDF is the inverse Fourier transform of the characteristic function, \ie
\bea
\label{eq:PDF:chi}
P\left(\zeta_{\mathrm{cg}}, \phi\right) = \frac{1}{2\pi} \int^{\infty}_{-\infty} \ee^{-it\left[\zeta_{\mathrm{cg}}+f_1\left(\phi\right)\right]} \chi_{\N}\left(t,\phi\right)\dd t\, ,
\eea
where we have used \Eq{eq:chideltaN:chiN}. The calculational programme is thus the following: solve \Eq{eq:ODE:chi} with boundary conditions~(\ref{eq:boundary:chi}), calculate $f_1$ either taking $n=1$ and $f_0=1$ in \Eq{eq:fn:generalsolution} or by noting that $f_1(\phi) = -i \partial\chi_\N /\partial t(t=0,\phi)$, calculate the PDF of curvature perturbations with \Eq{eq:PDF:chi}, and then the mass fraction of PBHs with \Eq{eq:beta:def}.
\subsubsection*{Example: quadratic potential}
\begin{figure}[t]
\begin{center}
\includegraphics[width=0.6\textwidth]{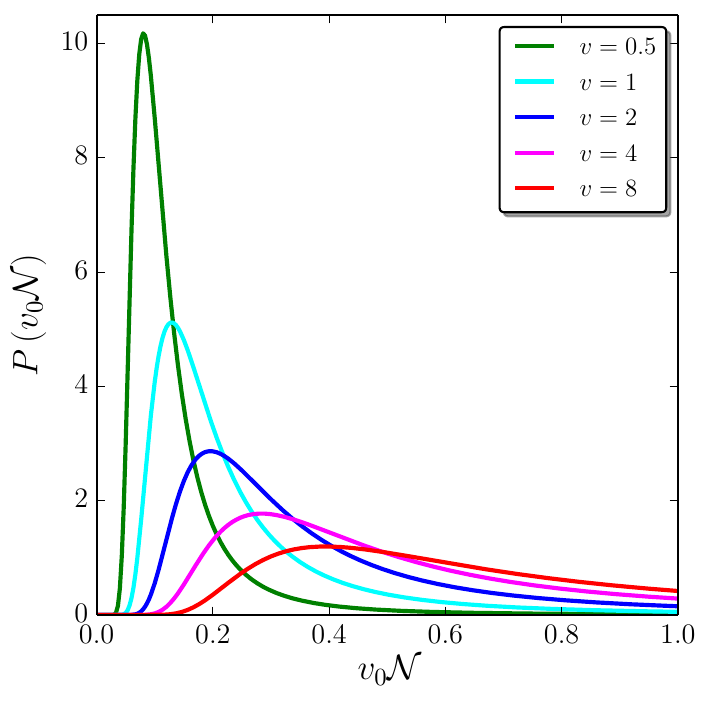}
\caption{Probability distributions of the number of \efolds~$\N$, rescaled by $v_0$, realised in the quadratic potential~(\ref{eq:quadratic:potential:example}) between an initial field value $\phi$ parametrised by $v(\phi)$ given in the legend, and $\phi_\uend = \sqrt{2}\Mp$ where inflation ends by slow-roll violation. The values displayed for $v$ correspond to very high energies far from the observational window of this model but this is for illustrative purpose only. When $v$ increases, one can see that the PDF has a larger mean value, a larger spread and seems to be less Gaussian, which motivates the need to go beyond Gaussian techniques.} 
\label{fig:pdf_quadratic_solution}
\end{center}
\end{figure}
In order to illustrate this computational programme, let us consider the case of a quadratic potential
\bea
\label{eq:quadratic:potential:example}
v(\phi) = v_{0}\left(\frac{\phi}{\Mp}\right)^2 \, .
\eea
In this case, \Eq{eq:ODE:chi} together with the boundary conditions~(\ref{eq:boundary:chi}) has an exact solution. Taking the $\phiuv \rightarrow \infty$ limit, it is given by 
\bea
\label{eq:chi:quadratic}
\chi_{\N}(t, \phi) =& \left[ \frac{v(\phi)}{v(\phiend)}\right]^{\frac{1 - \alpha(t)}{4}} \frac{{}_{1}F_{1} \left[ \frac{\alpha(t) - 1}{4} ; 1 + \frac{\alpha(t)}{2} ; -\frac{1}{v(\phi)} \right]}{{}_{1}F_{1} \left[ \frac{\alpha(t) - 1}{4} ; 1 + \frac{\alpha(t)}{2} ; -\frac{1}{v(\phiend)} \right]}\, ,
\eea
where $\alpha(t) = \sqrt{1 - \frac{4 i t}{v_{0}}}$ and ${}_1F_1(x;y;z)$ is the Kummer confluent hypergeometric function \cite{Olver:2010:NHM:1830479:hypergeometric, Abramovitz:1970aa:hypergeometric}\footnote{Note that these references use the notation $M(a,b,z)$ for the Kummer confluent hypergeometric function while we use the notation ${}_1F_1$.}. The inverse Fourier transform of \Eq{eq:chi:quadratic} can then be computed numerically, which gives rise to the PDF displayed in \Fig{fig:pdf_quadratic_solution}. At small $v(\phi)$ the PDF is rather peaked and almost Gaussian, while at large $v(\phi)$, it is more spread and deviates more from a Gaussian distribution. In \Secs{sec:ClassicalLimit} and~\ref{sec:StochasticLimit} we will study these two limits one by one, \ie the classical limit where the stochastic corrections are small and the PDF is almost Gaussian, and the stochastic limit where quantum diffusion dominates the inflaton dynamics.
\subsection{The heat equation approach}
\label{sec:heatequation}
Before investigating the classical and stochastic limits, let us note that the problem can be reformulated in terms of a heat equation for the PDF $P(\N,\phi)$. Indeed, if one plugs \Eq{eq:chi:P} into \Eq{eq:ODE:chi}, the two first terms apply on $P(\N,\phi)$ directly, while the third one is given by $i t \chi_\N = \int_{-\infty}^{\infty} \dd \N P(\N,\phi)\partial e^{it\N}/\partial\N = - \int_{-\infty}^{\infty} \dd \N e^{it\N} \partial P/\partial \N \left( \N, \phi \right)$. Here, in the first expression, we have simply differentiated \Eq{eq:chi:P} with respect to $\N$, and in the second expression, we have integrated by parts (the boundary terms vanish since $P(\N=\pm\infty,\phi)$ must vanish for the distribution to be normalisable). This gives rise to the heat equation
\bea
\label{eq:heat:phi}
\left(\frac{\partial^2}{\partial\phi^2}  - \frac{v^\prime}{v^2} \frac{\partial}{\partial\phi} - \frac{1}{v\Mp^2} \frac{\partial}{\partial\N}\right) P\left( \N, \phi \right) = 0\, .
\eea 
Instead of the infinite set of uncoupled differential equations given in \Eq{eq:ODE:chi}, the problem is now reformulated in terms of a single, but partial, differential equation. Let us note that \Eq{eq:heat:phi} does not have the structure of a Fokker-Planck equation, and should in any case not be confused with the usual Fokker-Planck equation considered in stochastic inflation which governs the PDF of the field value.\footnote{The Langevin equation~(\ref{eq:intro:Langevin}) gives rise to Fokker-Planck equation for the probability density $p(N,\phi )$ of the field to be at $\phi$ at time $N$, which, in the It\^{o} interpretation, is given by
\bea
\label{eq:FokkerPlanck}
\frac{\partial^2}{\partial\phi^2}\left[v p(N,\phi)\right] +\frac{\partial}{\partial\phi}\left[\frac{v'}{v}p(N,\phi)\right]-\frac{1}{\Mp^2}\frac{\partial}{\partial N}p(N,\phi)= 0\, .
\eea
This equation does not coincide with \Eq{eq:heat:phi} for $P(\N,\phi)$ which governs the probability to realise $\N$ \efolds~starting from $\phi$.} When plugging the boundary conditions~(\ref{eq:boundary:chi}) into \Eq{eq:chi:P}, one obtains
\bea
\label{eq:boundary:heat}
P\left(\N, \phiend\right) = \delta(\N)\, ,\quad
\frac{\partial P}{\partial \phi} \left(\N, \phiuv \right) = 0 \, .
\eea
These form the boundary conditions associated to \Eq{eq:heat:phi}.

In order to show that \Eq{eq:heat:phi} has the structure of a heat equation as announced above, one can introduce a change of field variable
\bea
u(\phi) = \int_{\phiend}^\phi \ee^{-\frac{1}{v(\tilde{\phi})}} \frac{\dd \tilde{\phi}}{\Mp}\, ,
\eea
which allows us to rewrite \Eq{eq:heat:phi} as
\bea
\label{eq:heat:u}
\left( v \ee^{-\frac{2}{v}}\frac{\partial^2}{\partial u^2}-\frac{\partial}{\partial \N}\right) P(\N,u) = 0 .
\eea
This is a heat equation for a one dimensional medium with diffusivity $v \ee^{-2/v}$, where $\N$ plays the role of time and $u$ the role of space. However, let us stress that heat equations are usually endowed with boundary conditions of a different type as from those in \Eq{eq:boundary:heat}, since in standard heat equations, one usually gives the spatial temperature distribution at an initial time, while \Eq{eq:boundary:heat} involves distributions of times at fixed spatial positions. This is why the numerical methods developed in the literature to solve heat equations would need to be adapted to this kind of boundary conditions but they may provide efficient ways to solve the problem at hand, \eg, in the context of multi-field inflation.
\section{Expansion about the classical limit}
\label{sec:ClassicalLimit}
In the limit of small quantum diffusion, the ``classical'' limit, one needs to check that our formulation allows one to recover the standard results recalled in \Sec{sec:Intro} around \Eq{eq:beta:erfc}. This is the goal of this section, where we also calculate the leading order deviation from the standard result in order to best determine its range of validity. From the heat equation~(\ref{eq:heat:u}), we saw that the diffusivity increases with $v$, which implies that the classical limit has $v\ll 1$ (this condition is not enough to define the classical regime as we will see below but it constitutes a fair starting point). We thus perform an expansion in increasing powers of $v$, first in the characteristic function approach introduced in \Sec{sec:CharacteristicFunction}, and then in the heat equation approach introduced in \Sec{sec:heatequation}. We will see that the former is much more convenient than the later which only yields limited results in the classical limit.
\subsection{The characteristic function approach}
\label{sec:classicalAppr:characteristicMethod}
In the ordinary differential equation satisfied by the characteristic function, \Eq{eq:ODE:chi}, an expansion in $v$ is equivalent to an expansion in the diffusion term, involving $\partial^2/\partial\phi^2$.
\subsubsection{Leading order}
\label{sec:classicalAppr:characteristicMethod:LO}
At leading order (LO) in the classical limit, the diffusion term in \Eq{eq:ODE:chi} can be simply neglected, and one has
\bea
\label{eq:ODE:chi:classical}
\left(- \frac{v^\prime}{v} \frac{\partial}{\partial\phi} + \frac{it}{\Mp^2}\right)\chi^{\lo}_\N(t,\phi)=0\, .
\eea 
Making use of the first boundary condition in \Eq{eq:boundary:chi},\footnote{In the expansion about the classical limit, the second boundary condition in \Eq{eq:boundary:chi} cannot be satisfied simultaneously with the first condition. This is why the solutions presented here are, strictly speaking, only valid in the limit $\phiuv \rightarrow \infty$.} this equation can be solved as
\bea
\label{eq:chi:classical:LO}
\chi_\N^\mathrm{\lo}(t,\phi) = \exp\left[it\int_{\phiend}^{\phi} \frac{v(x)}{\Mp^2v^\prime(x)}\dd x\right]\, .
\eea
Note that the integral in the argument of the exponential is the classical number of \efolds, which is also the mean number of \efolds~at leading order in the classical limit, \ie the leading order saddle point expansion of \Eq{eq:fn:generalsolution}~\cite{Vennin:2016wnk},
\bea
\label{eq:f1lo}
 f_1^\lo(\phi)=\frac{1}{\Mp^2}\int_{\phiend}^\phi  \frac{v(x)}{v^\prime(x)} \dd x\, .
\eea 
This is consistent with the formula given below \Eq{eq:PDF:chi}, namely $f_1(\phi) = -i \partial\chi_\N /\partial t(t=0,\phi)$.  As a consequence, \Eq{eq:chideltaN:chiN} implies that $\chi_{\delta N_{\mathrm{cg}}} = 1$, and hence its inverse Fourier transform is $P^\lo\left(\delta N_\mathrm{cg}, \phi\right) = \delta \left(\delta N_\mathrm{cg}\right)$, \ie a Dirac distribution centred around $\delta N_\mathrm{cg} = 0$. 
Thus, at leading order in the classical limit, one simply shuts down quantum diffusion, the dynamics are purely deterministic, $\delta\N \equiv 0$ and there are no curvature perturbations. 
\subsubsection{Next-to-leading order}
\label{sec:nlo}
One thus needs to go to next-to-leading order (NLO) to incorporate curvature perturbations. At NLO, the LO solution~(\ref{eq:chi:classical:LO}) can be used to evaluate the term $\chi^{-1}\partial^2\chi/\partial\phi^2$ in \Eq{eq:ODE:chi}, which then becomes
\bea
\frac{\partial}{\partial\phi}\chi_\N^\nlo -\frac{v^2}{v^\prime} \left(\frac{it}{v\Mp^2}+ \frac{1}{\chi_\N^\lo}\frac{\partial^2 \chi_\N^\lo}{\partial\phi^2} \right)
\chi_\mathcal{N}^\nlo  =0\, .
\eea
Making use of the first boundary condition in \Eq{eq:boundary:chi}, the solution of this first order ordinary differential equation is
\bea
\label{eq:chi:classical:interativeSolution}
\chi_\N^\nlo (t,\phi) = \exp\left\lbrace \int_{\phi_\uend}^\phi \left[ \frac{itv(x)}{\Mp^2v' (x)} + \frac{v^2(x)}{v'(x)} \frac{1}{\chi_\N^\lo(x)}\frac{\partial^2\chi_\N^\lo}{\partial\phi^2}(x) \right] \dd x\right\rbrace\, .
\eea
Notice that if ones replaces $\lo$  by an arbitrary n${}^\mathrm{th}$ order and $\nlo$ by the n+1${}^\mathrm{th}$ order of the classical expansion, this equation is valid at any order since it is nothing but the iterative solution of \Eq{eq:ODE:chi}. At NLO, plugging \Eq{eq:chi:classical:LO} into \Eq{eq:chi:classical:interativeSolution}, one obtains
\bea
\label{eq:chi:classical:NLO}
\chi_{\N}^\nlo(t,\phi)  = \exp\left[ it f_1^\nlo\left(\phi\right) - \gamma_{1}^\nlo v t^2  \right] \, ,
\eea
where $f_1^\nlo$ is the mean number of \efolds~at NLO~\cite{Vennin:2016wnk},
\bea
f_1^\nlo(\phi)=\frac{1}{\Mp^2}\int_{\phiend}^\phi \dd x \left(\frac{v}{v^\prime} +\frac{v^2}{{v^\prime}}-\frac{v^3 v^{\prime\prime}}{{v^\prime}^3}\right)\, ,
\eea
and we have defined
\bea
\label{eq:gamma1:nlo:def}
\gamma_{1}^\nlo = \frac{1}{v\Mp^4} \int_{\phiend}^\phi \dd x \frac{v^4}{{v^\prime}^3} \, .
\eea
From this expression, \Eq{eq:chideltaN:chiN} implies that $\chi_{\delta N_\mathrm{cg}}^\nlo\left(t,\phi\right) = \ee^{-\gamma_{1}^\nlo v t^2}$, that is to say $\chi_{\delta\N}^\nlo$ is a Gaussian and hence its inverse Fourier transform $P^\nlo\left(\zeta_\mathrm{cg}, \phi\right)$ is also a Gaussian and is given by
\bea
\label{eq:PDF:classical:NLO}
P^\nlo(\zeta_{\mathrm{cg}}, \phi) = \frac{1}{\sqrt{4\pi\gamma_{1}^\nlo v}} \exp\left(-\frac{\zeta_{\mathrm{cg}}^2}{4\gamma_{1}^\nlo v}\right) \, .
\eea
A crucial remark is that at this order, the power spectrum~(\ref{eq:Pzeta:stochaDeltaN}) is given by~\cite{Vennin:2015hra} \Eq{eq:classicalPS}, so that the variance of the Gaussian distribution~(\ref{eq:PDF:classical:NLO}) reads $2\gamma_{1}^\nlo v = \int_{\phiend}^{\phi} \calP_\zeta^{\nlo} f_1^{'\lo} \dd x$. This precisely matches the standard result recalled above \Eq{eq:beta:erfc}, namely that $P(\zeta_{\mathrm{cg}})$ is a Gaussian PDF with standard deviation given by the integrated power spectrum $\left\langle \zeta_{\mathrm{cg}}^2 \right\rangle = \int_{k}^{k_\uend} \calP_\zeta(\tilde{k})\dd \ln \tilde{k}$, since $\dd \ln k \simeq \dd N = f_1'(\phi) \dd \phi$ at leading order in slow roll.
\subsubsection{Next-to-next-to-leading order}
\label{sec:classical:nnlo}
\begin{figure}[t]
\begin{center}
\includegraphics[width=0.6\textwidth]{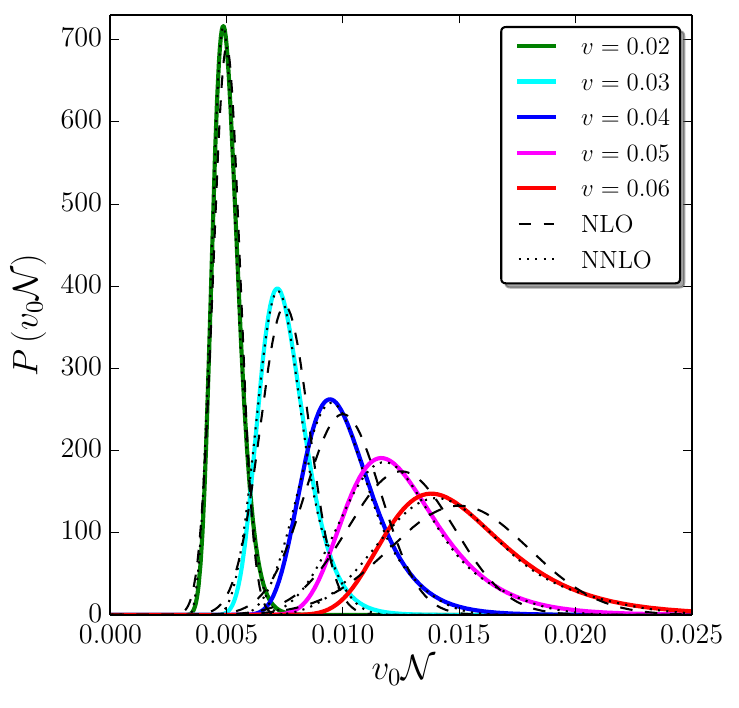}
\caption{Probability distributions of the number of \efolds~$\N$, rescaled by $v_0$, realised in the quadratic potential~(\ref{eq:quadratic:potential:example}) between an initial field value $\phi$ parametrised by $v(\phi)$ given in the legend, and $\phi_\uend = \sqrt{2}\Mp$ where inflation ends by slow-roll violation, as in \Fig{fig:pdf_quadratic_solution}. The black dashed lines correspond to the NLO (Gaussian) approximation~(\ref{eq:PDF:classical:NLO}), while the dotted lines stand for the NNLO approximation~(\ref{eq:PDF:NNLO}). The smaller $v$ is, the better these approximations are, and the NNLO approximation is substantially better than the NLO one.} 
\label{fig:pdf_quadratic_solution_nnlo}
\end{center}
\end{figure}

In order to study the first non-Gaussian corrections to the standard result, one needs to go to next-to-next-to-leading order (NNLO). As explained in \Sec{sec:nlo}, one can simply increment the order of the iterative relation~(\ref{eq:chi:classical:interativeSolution}), \ie replace $\lo$ by $\nlo$ and $\nlo$ by $\nnlo$. Plugging in \Eq{eq:chi:classical:NLO}, and making use of \Eq{eq:chideltaN:chiN}, this gives rise to
\bea
\label{eq:chiNNLO}
\chi_{\delta N_{\mathrm{cg}}}^\nnlo\left(t,\phi\right)  = \exp\left( - \gamma_{1}^\nnlo v t^2 - i\gamma_{2}^\nnlo v^2t^3 
 \right)  \, ,
\eea
where we have only kept the terms that are consistent at that order and where we have defined
\bea
\label{eq:gamma:nnlo:def}
\gamma_{1}^\nnlo &= \frac{1}{v\Mp^4 }  \int_{\phiend}^\phi\dd x\left(\frac{v^4 }{v'^3}+6\frac{v^5 }{v'^3}-5\frac{v^6 v''}{v'^5}\right) \, ,\\
\gamma_{2}^\nnlo &= \frac{2}{v^2 \Mp^6}  \int_{\phiend}^\phi\dd x \frac{v^7}{v'^5} \, .
\eea
One can already see that since the characteristic function is not a Gaussian, the PDF is not a Gaussian distribution. Using \Eq{eq:PDF:chi}, it is given by
\bea
\label{eq:PDF:NNLO:chi}
P^{\nnlo}\left( \delta N_{\mathrm{cg}}, \phi \right) = \frac{1}{2\pi}\int_{-\infty}^{\infty} \dd t \exp\left( -it\delta N_{\mathrm{cg}} - \gamma_{1}^\nnlo vt^2 + i\gamma_{2}^\nnlo v^2t^3  \right) \, .
\eea
In this integral, the second term in the argument of the exponential makes the integrand become negligible when $\gamma_1^\nnlo v t^2 \gg 1$, \ie for $\vert t \vert \gg t_\uc$ where $t_\uc = (\gamma_{1}^\nnlo v )^{-1/2}$. When $t=\pm t_\uc$, the ratio between the third and the second terms in the argument of the exponential of \Eq{eq:PDF:NNLO:chi} is of order $(\gamma_2^\nnlo / \gamma_1^\nnlo) \sqrt{v/\gamma_1^\nnlo}$, \ie of order $\sqrt{v}$ in an expansion in $v$ since the $\gamma_i$ parameters have been defined to carry no dimension of $v$ (at least at their leading orders). This is why, over the domain of integration where most of the contribution to the integral comes from, the third term is negligible and can be Taylor expanded. One obtains 
\bea
\label{eq:PDF:NNLO}
P^{\nnlo}\left( \zeta_\mathrm{cg}, \phi \right)  = \frac{1}{\sqrt{4 \pi \gamma_{1}^\nnlo v}} \exp\left(-\frac{\zeta_\mathrm{cg}^2}{4\gamma_{1}^\nnlo v}\right)\left[ 1 - \frac{\gamma_{2}^\nnlo}{8\left({\gamma_{1}^\nnlo}\right)^3 v} \zeta_\mathrm{cg} \left( 6 \gamma_{1}^\nnlo v - \zeta_\mathrm{cg}^2 \right) \right] \, .
\eea 
For the quadratic potential example discussed in \Sec{sec:CharacteristicFunction}, in \Fig{fig:pdf_quadratic_solution_nnlo} we have reproduced \Fig{fig:pdf_quadratic_solution} (for different values of $v$ to better illustrate the behaviours of the classical approximations) where we have superimposed the NLO approximation~(\ref{eq:PDF:classical:NLO}) and the NNLO approximation~(\ref{eq:PDF:NNLO}). One can check that these approximations become better at smaller values of $v$ as expected, and that the NNLO approximation always provides a better fit than the NLO one. 

As a consistency check, one can verify that the distribution~(\ref{eq:PDF:NNLO}) yields the same moments at NNLO order as the ones derived in \Ref{Vennin:2015hra} by calculating the integrals~(\ref{eq:fn:generalsolution}) with a saddle-point approximation technique at NNLO. For the second moment, one has $\langle \delta N_{\mathrm{cg}}^2 \rangle = \int^{\infty}_{-\infty} \zeta_{\mathrm{cg}}^2 P^{\nnlo} ( \zeta_\mathrm{cg}, \phi ) \dd \zeta_{\mathrm{cg}} = 2 \gamma_1^\nnlo v$, which coincides with Eq.~(3.35) of \Ref{Vennin:2015hra}. Similarly for the third moment, $\langle \delta N_{\mathrm{cg}}^3 \rangle = \int^{\infty}_{-\infty} \zeta_{\mathrm{cg}}^3 P^{\nnlo} ( \zeta_\mathrm{cg}, \phi ) \dd \zeta_{\mathrm{cg}} = 6 \gamma_2^\nnlo v^2$, which coincides with Eq.~(3.37) of \Ref{Vennin:2015hra}. The two methods, \ie the iterative solution~(\ref{eq:chi:classical:interativeSolution}) of the characteristic function equation and the saddle-point expansion of the integrals~(\ref{eq:fn:generalsolution}), are therefore equivalent.

Let us also note that the characteristic function, $\chi_{\cal N}(t,\phi)$ defined in \Eq{eq:characteristicFunction:def}, is closely related to the cumulant generating function for the probability distribution
\bea
\label{defKtau}
K_{\cal N}(\tau,\phi) = \ln \langle e^{\tau {\cal N}(\phi)} \rangle = \sum_{n=1}^\infty \frac{\kappa_n(\phi)}{n!} \tau^n \,.
\eea
By comparing \Eqs{eq:characteristicFunction:def} and~(\ref{defKtau}) indeed, one simply has $\chi_{\cal N}(t,\phi)=\exp\left[K_{\cal N}(it,\phi)\right]$. If we now compare \Eqs{eq:chiNNLO} and~(\ref{defKtau}), we can read off the first cumulants
\bea
\kappa_2(\phi) = 2 v \gamma_{1} \,,  \quad \kappa_3(\phi) = 6v^2 \gamma_{2} \,.
\eea
One measure of the deviation from a Gaussian distribution is the skewness of the distribution which is determined by the ratio of these cumulants
\bea
\label{eq:gamma_skew}
\gamma_{\mathrm{skew}} \equiv \frac{\kappa_3}{\kappa_2^{3/2}} = \frac{3v^{1/2}\gamma_{2}}{\sqrt{2}(\gamma_{1})^{3/2}}
\,.
\eea
Since $\gamma_2$ is non-vanishing at next-to-next-to-leading (and higher) order only, the NNLO term thus represents the first non-Gaussian correction to the standard Gaussian result obtained at NLO in the classical limit. 

At this order, the distribution is positively skewed, which is indeed the case for all the distributions displayed in \Figs{fig:pdf_quadratic_solution} and~\ref{fig:pdf_quadratic_solution_nnlo}. One can also note that the parameter introduced below \Eq{eq:PDF:NNLO:chi}, that must be small in order for  the classical expansion to be valid at NNLO, exactly coincides with $\gamma_{\mathrm{skew}}$. The above formulae are therefore correct in the limit $\gamma_{\mathrm{skew}}\ll 1$ only. Finally, the correcting term in the brackets of \Eq{eq:PDF:NNLO} can be expressed as $\gamma_{\mathrm{skew}} \left(\zeta_\mathrm{cg}^2/\left\langle \zeta_\mathrm{cg}^2 \right\rangle\right)^{3/2}$, where we have used the relation $\langle \delta N_{\mathrm{cg}}^2 \rangle =  2 \gamma_1^\nnlo v$ given above together with \Eq{eq:gamma_skew}. This shows that $\gamma_{\mathrm{skew}}\ll 1$ only ensures the correcting term to be small when $\zeta_\mathrm{cg}^2$ is of order $\left\langle \zeta_\mathrm{cg}^2 \right\rangle$, \ie around the maximum of the distribution. The classical approximation is therefore an expansion about the maximum of the distribution that should be expected to fail in the tail. Since the PBH threshold $\zeta_\uc$ is usually in the far tail of the distribution (in the standard calculation recalled in \Sec{sec:Intro}, at the level of the observational bounds, one has $\left\langle \zeta_\mathrm{cg}^2 \right\rangle \sim 10^{-2} \ll \zeta_\uc^2\sim 1$), one may need to go beyond the classical approximation in such cases.
\subsection{The heat equation approach}
Before moving on to the stochastic limit, let us briefly explain how the heat equation approach proceeds in the classical limit. At LO, neglecting the diffusion term in \Eq{eq:heat:phi}, one has to solve $\Mp^2 v'/v \, \partial P/\partial \phi + \partial P/\partial \N =0$, with the first boundary condition of \Eq{eq:boundary:heat}. Using the method of characteristics to solve first-order partial differential equations, one obtains
\bea
\label{eq:heat:classical:LO:solution}
P^{\lo} \left( \N, \phi \right) = \delta \left[ \N - f_1^{\lo}\left(\phi\right) \right] \, ,
\eea
where $ f_1^{\lo}$ has been defined in \Eq{eq:f1lo} and corresponds to the classical number of \efolds, which is also the mean number of e-folds at leading order in the classical limit. One therefore recovers the result of \Sec{sec:classicalAppr:characteristicMethod:LO}. At NLO, one can use \Eq{eq:heat:classical:LO:solution} to calculate the diffusive term in \Eq{eq:heat:phi} and iterate the procedure. However, by doing so, one has to solve a first-order partial differential equation with a source term that involves derivatives of the Dirac distribution. This makes the solving procedure technically complicated, and we therefore do not pursue this direction further since a simpler way to obtain the solution was already presented in \Sec{sec:classicalAppr:characteristicMethod}. One can already see the benefit of having two solving procedures at hand, which will become even more obvious in what follows.
\section{The stochastic limit}
\label{sec:StochasticLimit}
\begin{figure}[t]
\begin{center}
\includegraphics[width=0.6\textwidth]{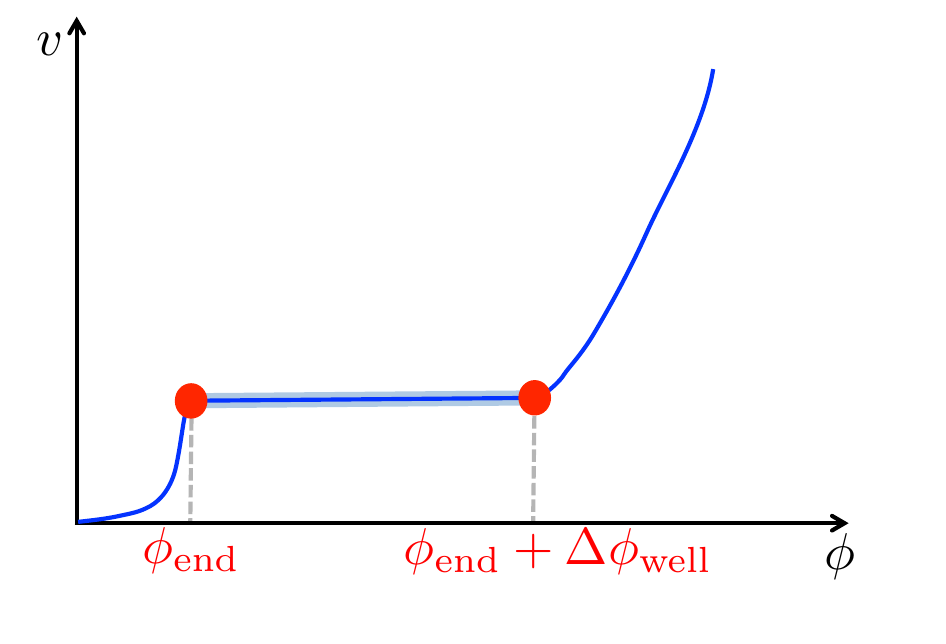}
\caption{
Schematic representation of the single-field stochastic dynamics solved in \Sec{sec:StochasticLimit}, where the potential may be taken to be 
exactly constant over the ``quantum well'' regime delimited by $\phi_\uend$ and $\phi_\uend+\Delta\phiwell$. Inflation terminates at $\phi_\uend$, where either the potential becomes very steep or a mechanism other than slow-roll violation ends inflation, and a reflective wall is placed at $\phi_\uend+\Delta\phiwell$, which can be seen as the point where the dynamics become classically dominated and the classical drift prevents the field from escaping the quantum well.}  
\label{fig:sketch2}
\end{center}
\end{figure}
We now consider the opposite limit where the inflaton field dynamics are dominated by quantum diffusion. This is the case if the potential is exactly flat, since then the slow-roll classical drift vanishes. We thus consider a potential that is constant between the two values $\phi_\uend$ and $\phi_\uend+\Delta\phiwell$, where $\Delta\phiwell$ denotes the width of this ``quantum well''. Inflation terminates when the field reaches $\phi_\uend$ (where either the potential is assumed to become very steep, or a mechanism other than slow-roll violation must be invoked to end inflation), and a reflective wall is located at  $\phi_\uend+\Delta\phiwell$, which can be seen as the point where the dynamics become classically dominated so that the probability for field trajectories to climb up this part of the potential and escape the quantum well can be neglected. The situation is depicted in \Fig{fig:sketch2}, and in \Sec{sec:PBH} we will see why these assumptions allow one to study most cases of interest. 
\subsection{The characteristic function approach}
If the potential $v=v_0$ is constant, the potential gradient term vanishes in \Eq{eq:ODE:chi} and making use of the boundary conditions~(\ref{eq:boundary:chi}), where $\phiuv$ is replaced by $\phi_\uend+\Delta\phiwell$, one obtains
\bea
\label{eq:chiN:cosh}
\chi_\N\left(t, \phi \right) = \frac{\cosh\left[\alpha \sqrt{t} \mu \left(x-1\right)\right]}{\cosh\left(\alpha \sqrt{t} \mu\right)}\, .
\eea
In this expression, $x\equiv (\phi-\phi_\uend)/\Delta\phiwell$, $\alpha\equiv (i-1)/\sqrt{2}$, and we have introduced the parameter
\bea
\label{eq:def:mu}
\mu^{2} = \frac{\Delta\phiwell^2}{v_{0}\Mp^2} 
\eea
which is the ratio between the squared width of the quantum well and its height, in Planck mass units, and which is the only combination through which these two quantities appear.

\begin{figure}[t]
\begin{center}
\includegraphics[width=0.6\textwidth]{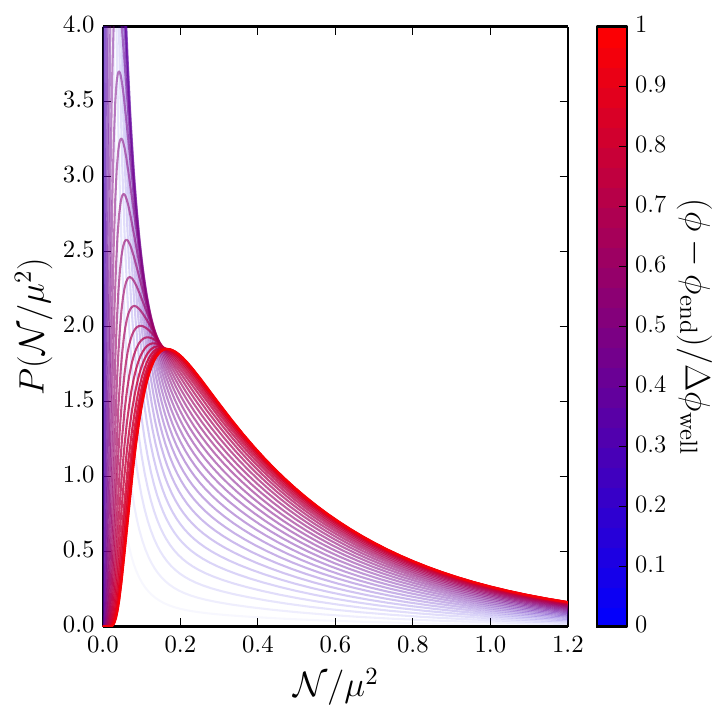}
\caption{Probability distributions of the number of \efolds~$\N$, rescaled by $\mu^2$, realised in the constant potential depicted in \Fig{fig:sketch2} between $\phi$ and $\phi_\uend$, where different colours correspond to different values of $\phi$. When $\phi$ approaches $\phi_\uend$, the distribution becomes more peaked and the transparency of the curves is increased for displayed purposes.} 
\label{fig:pdf_stochastic}
\end{center}
\end{figure}

The PDF can be obtained by inverse Fourier transforming \Eq{eq:chiN:cosh}, see \Eq{eq:PDF:chi}, which can be done after Taylor expanding the characteristic function~(\ref{eq:chiN:cosh}) and inverse Fourier transforming each term in the sum. This leads to
\bea
\label{eq:stocha:CharacteristicFunctionMethod:PDF}
& P\left(\N, \phi \right) = \frac{1}{2\sqrt{\pi}}\frac{\mu}{\N^{3/2}} \times \\
& \quad \left\lbrace
 \sum_{n=0}^\infty (-1)^n \left[2(n+1) - x \right]
 \ee^{- \frac{\mu^2}{4\N} \left[2(n+1) - x \right]^2} 
+ \sum_{n=0}^\infty (-1)^n \left[2n + x \right]
 \ee^{- \frac{\mu^2}{4\N} \left[2n + x \right]^2} 
\right\rbrace\, .
\eea
This PDF is displayed in \Fig{fig:pdf_stochastic} for different values of $x$. Interestingly, \Eq{eq:stocha:CharacteristicFunctionMethod:PDF} can be resummed to give a closed form when combined with the result from the heat equation approach presented below in \Sec{sec:StochasticLimit:HeatEquationApproach}. For now, we can derive closed form expressions at both boundaries of the quantum well, \ie in the two limits $\phi\simeq \phi_\uend+\Delta\phiwell$ and $\phi\simeq\phiend$.
\subsubsection*{Reflective boundary of the quantum well}
In the case where $\phi =  \phi_\uend+\Delta\phiwell$, or $x=1$, \ie at the reflective boundary of the quantum well, \Eq{eq:stocha:CharacteristicFunctionMethod:PDF} reduces to $P(\N, \phiwell) =\mu/\sqrt{\pi} \N^{-3/2}  \sum_{n=0}^\infty (-1)^n (2n+1) \ee^{- \frac{\mu^2}{4\N}  (2n+1)^2}$. Making use of the elliptic theta functions~\cite{Olver:2010:NHM:1830479:theta, Abramovitz:1970aa:theta} introduced in \App{appendix:identities}, this can be rewritten as 
\bea
\label{eq:PDF:phiwall:thetaElliptic}
P\left(\N, \phi =  \phi_\uend+\Delta\phiwell\right) = \frac{\mu}{2 \sqrt{\pi} \N^{3/2}}  
\vartheta^\prime_1\left(0,\ee^{-\frac{\mu^2}{\N}}\right) \, ,
\eea
where $\vartheta_1^\prime$ is the derivative (with respect to the first argument) of the first elliptic theta function, see \Eq{eq:theta1prime:def}.
\subsubsection*{Absorbing boundary of the quantum well}
\begin{figure}[t]
\begin{center}
\includegraphics[width=0.496\textwidth]{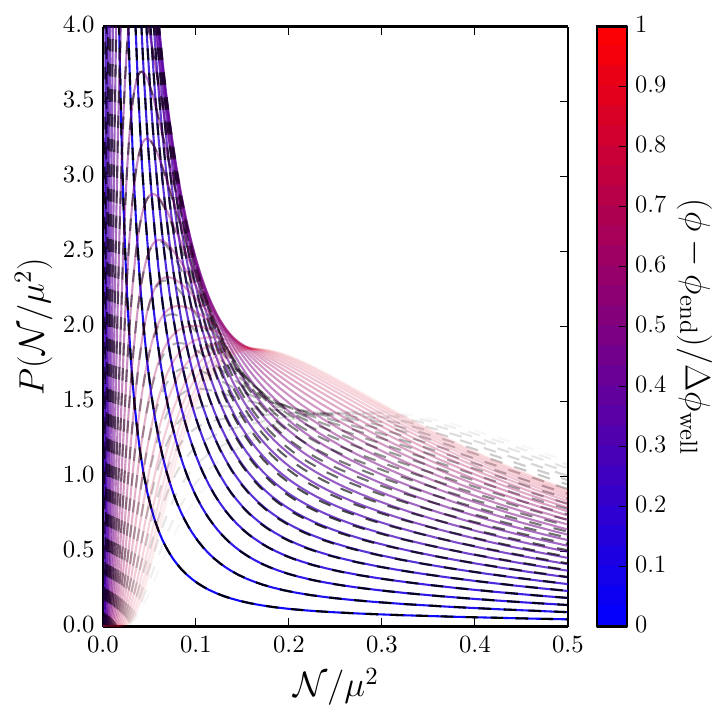}
\includegraphics[width=0.496\textwidth]{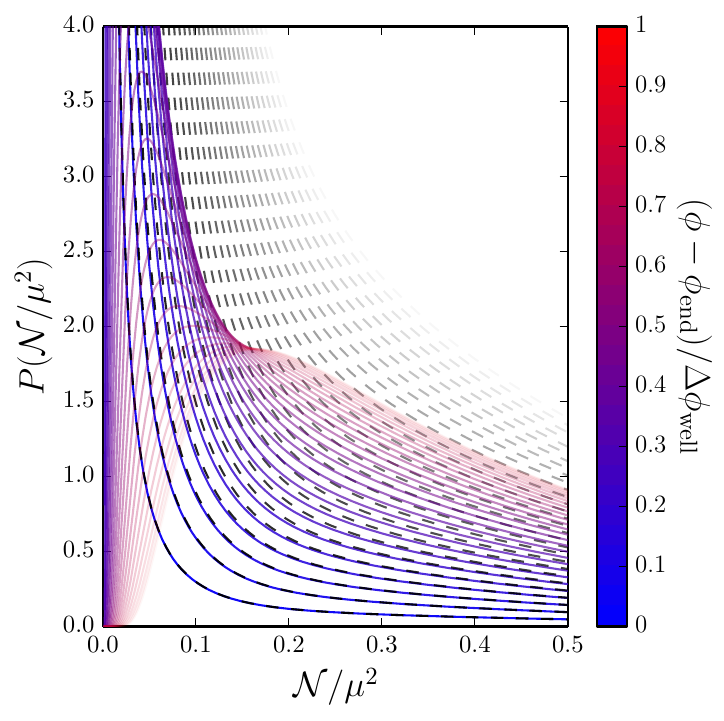}
\caption{Probability distributions of the number of \efolds~$\N$, rescaled by $\mu^2$, realised in the constant potential depicted in \Fig{fig:sketch2} between $\phi$ and $\phi_\uend$. In both panels, different colours correspond to different values of $\phi$, and the black dashed lines correspond to approximations. Left panel: the approximation~(\ref{eq:PDF:stocha:phiend:appr}) is displayed with the black dashed lines. Right panel: the approximation~(\ref{eq:PDF:stocha:phiend:appr:2}) is displayed with the black dashed lines. These approximations are valid close to the absorbing boundary of the quantum well where inflation ends. When $\phi$ increases, the approximation becomes worse, and the transparency of the curves is increased for displayed purposes, but one can see that the approximation~(\ref{eq:PDF:stocha:phiend:appr}) is excellent up to $(\phi-\phiend)/\Delta\phiwell \sim 0.3$, and slightly better than the approximation~(\ref{eq:PDF:stocha:phiend:appr:2}).} 
\label{fig:pdf_stochastic_appr}
\end{center}
\end{figure}

In the case where $\phi \simeq  \phi_\uend$, or $x\ll 1$, \ie at the absorbing boundary of the quantum well, an approximated formula can be obtained by noting that \Eq{eq:stocha:CharacteristicFunctionMethod:PDF} can be rewritten as $P(\N, \phi) = \mu /(2\sqrt{\pi}\N^{3/2})[x \ee^{-\mu^2 x^2/(4\N)}+F(-x)-F(x)]$, with $F(x)\equiv \sum_{n=0}^\infty (-1)^n [2(n +1)+ x]
 \ee^{- \frac{\mu^2}{4 \N} [2(n+1)+x]^2}$. In the limit where $x\ll 1$, $F(-x)-F(x)\simeq -2 x F'(0)$, where $F'(0)=1/2 - 1/2 \vartheta_4(0,\ee^{-\mu^2/\N})-\mu^2/(4\N)\vartheta_4''(0,\ee^{-\mu^2/4})$, see \Eq{eq:theta4primeprime:def}. This gives rise to
\bea
\label{eq:PDF:stocha:phiend:appr}
P\left(\N, \phi \simeq \phi_\uend \right) \simeq \frac{\mu x}{2\sqrt{\pi}\N^{3/2}}
\left[ \ee^{- \frac{\mu^2 x^2}{4 \N }} 
- 1 + \vartheta_4\left(0,\ee^{- \frac{\mu^2}{\N} } \right) + \frac{\mu^2}{2 \N} \vartheta_4^{\prime\prime}\left(0,\ee^{- \frac{\mu^2}{\N} } \right) 
\right] \, .
\eea
This approximation is superimposed to the full result~(\ref{eq:PDF:phiwall:thetaElliptic}) in the left panel of \Fig{fig:pdf_stochastic_appr}, where one can check that the agreement is excellent even up to $x\sim 0.3$.
\subsection{The heat equation approach}
\label{sec:StochasticLimit:HeatEquationApproach}
Let us now move on to the heat equation approach, since combined with the results of the characteristic function approach, this will allow us to derive a closed form for the PDF at arbitrary values of $x$. In the case of a constant potential, the heat equation~(\ref{eq:heat:phi}) becomes $(v_0\Mp^2\partial^2/\partial\phi^2 - \partial/\partial \N)P(\N, \phi ) = 0$. The second boundary condition of \Eq{eq:boundary:heat},  $\partial P/\partial\phi (\N, \phi_\uend+\Delta\phiwell) = 0$, leads to a Fourrier decomposition of the form
\bea
\label{eq:stochastic:heat:FourierDecomposition}
P\left(\N, \phi \right) = \sum_{n=0}^\infty \left\lbrace A_n\left(\N\right) \sin\left[\left(\frac{\pi}{2}+ n\pi\right)x\right]
+B_n\left(\N\right) \cos\left(n\pi x \right) \right\rbrace \, ,
\eea
where, by plugging \Eq{eq:stochastic:heat:FourierDecomposition} into the heat equation~(\ref{eq:heat:phi}), the coefficients $A_n$ and $B_n$ must satisfy
\bea
\frac{\partial A_n}{\partial \mathcal{N}} = -\frac{\pi^2}{\mu^2} \left(n+\frac{1}{2}\right)^2 A_n\, , \quad
\frac{\partial B_n}{\partial \mathcal{N}} = -\frac{\pi^2}{\mu^2} n^2 B_n\, .
\eea
This leads to
\bea
A_n(\mathcal{N})  = a_n \exp\left[ -\frac{\pi^2}{\mu^2} \left(n+\frac{1}{2}\right)^2 \mathcal{N}\right]\, , \quad
B_n(\mathcal{N})  = b_n  \exp\left(-\frac{\pi^2}{\mu^2} n^2 \mathcal{N} \right)\, ,
\eea
where $a_n$ and $b_n$ are coefficients that depend only on $n$. They can be calculated by identifying \Eqs{eq:stocha:CharacteristicFunctionMethod:PDF} and~(\ref{eq:stochastic:heat:FourierDecomposition}) in the $\N \to 0$ limit. In this limit, in \Eq{eq:stocha:CharacteristicFunctionMethod:PDF}, the term with $n=0$ of the second sum is the dominant contribution, and using the fact that $\ee^{-x^2/(4\sigma)}/ (2\sqrt{\pi \sigma}) \to \delta(x)$ when $\sigma\to 0$, hence $- x\ee^{-x^2/(4\sigma)}/ (4\sigma \sqrt{\pi \sigma}) \to \delta'(x)$ when $\sigma\to 0$, one has 
\bea
\label{eq:stochastic:PDF:Neq0}
P(\N,\phi)\underset{\N \to 0}{\longrightarrow} -2 v_0 \Mp^2 \delta'(\phi-\phi_\uend)\, .
\eea
In passing, one notes that this expression implies that $P(\N=0,\phi)=0$ when $\phi \neq \phi_\uend$, which is consistent with the continuity of the distribution when $\N=0$ and with the fact that the probability to realise a negative number of \efolds~obviously vanishes. The case $\phi=\phiend$ is singular because of the first boundary condition of \Eq{eq:boundary:heat}, which explains the singularity in \Eq{eq:stochastic:PDF:Neq0}. The coefficients $a_n$ and $b_n$ can then be expressed as $a_n = \int_{-1}^1 \dd x P(\N=0,\phi) \sin[(n+1/2)\pi x]$ for $n\geq 0$ and $b_n = \int_{-1}^1 \dd x P(\N=0,\phi) \cos[n\pi x]$ for $n\geq 0$, where we recall that the link between $\phi$ and $x$ is given above \Eq{eq:def:mu}. This gives rise to $a_n=2\pi(n+1/2)/\mu^2$ and $b_n=0$, hence 
\bea
\label{eq:stocha:HeatMethod:PDF:expansion}
P\left(\N, \phi \right) = & \frac{2 \pi}{\mu^2} \sum_{n=0}^\infty \left( n + \frac{1}{2} \right) \exp\left[ -\frac{\pi^2}{\mu^2} \left(n+\frac{1}{2}\right)^2 \N\right] \sin\left[x \pi\left(n+\frac{1}{2}\right)\right] \, ,
\eea
which can be written as
\bea
\label{eq:PDF:thetatwo}
P\left( \N, \phi \right) = & -\frac{\pi}{2 \mu^2} \vartheta_{2}' \left( \frac{\pi}{2}x, \ee^{-\frac{\pi^2}{\mu^2} \N} \right) \, ,
\eea
see \Eq{eq:theta2prime:def}. A few comments are in order.

First, let us stress that the results from both methods, the characteristic function one and the heat equation one, have been necessary to derive this closed form, since the expression coming from the characteristic function has allowed us to calculate the coefficients $a_n$ and $b_n$ in the heat equation solution. This further illustrates how useful it is to have two approaches at hand.

Second, the expansion~(\ref{eq:stocha:HeatMethod:PDF:expansion}) is an alternative to the one given in \Eq{eq:stocha:CharacteristicFunctionMethod:PDF} for the PDF. One can numerically check that they are identical, and in \Fig{fig:pdf_stochastic}, $P(\N,\phi)$ is displayed as a function of $\N$ for various values of $\phi$. The difference between \Eqs{eq:stocha:CharacteristicFunctionMethod:PDF} and~(\ref{eq:stocha:HeatMethod:PDF:expansion}) is that they correspond to expansions around different regions of the PDF. In \Eq{eq:stocha:CharacteristicFunctionMethod:PDF}, since one is summing over increasing powers of $\ee^{-1/\N}$, one is expanding around $\N=0$, \ie on the ``left'' tail of the distribution. In \Eq{eq:stocha:HeatMethod:PDF:expansion} however, since one is summing over increasing powers of $\ee^{-\N}$, one is expanding around $\N=\infty$, \ie on the ``right'' tail of the distribution. Therefore, if one wants to study the PDF by truncating the expansion at some fixed order $n$, one should choose to work with the expression that better describes the part of the distribution one is interested in, so that both expressions can a priori be useful (let us stress again that, in the limit where all terms in the sums are included, both expressions match exactly for all values of $\N$). 

Third,  by plugging $x=1$ in \Eq{eq:PDF:thetatwo}, one obtains an expression for $P(\N,\phi=\phiend+\Delta\phiwell)$ that is an alternative to \Eq{eq:PDF:phiwall:thetaElliptic} even if both formulae involve elliptic theta functions. In \App{appendix:identities}, we show that both expressions are equivalent, due to identities satisfied by the elliptic theta functions. In fact, a third expression for $P(\N,\phi=\phiend+\Delta\phiwell)$ can even be obtained by plugging $x=1$ into \Eq{eq:stocha:HeatMethod:PDF:expansion} and the consistency with the two other ones is also shown in \App{appendix:identities}.

Fourth, an approximated formula for the PDF in the limit $\phi\sim\phiend$ can be derived by Taylor expanding \Eq{eq:PDF:thetatwo}, 
\bea
\label{eq:PDF:stocha:phiend:appr:2}
P\left(\N, \phi \simeq \phiend \right) \simeq -\frac{\pi^2}{4 \mu^2} x \vartheta_{2}'' \left(0, \ee^{-\frac{\pi^2}{\mu^2}\N} \right )\, ,
\eea
see \Eq{eq:theta2primeprime:def}. This provides an alternative to the approximation~(\ref{eq:PDF:stocha:phiend:appr}), that is displayed in the right panel of \Fig{fig:pdf_stochastic_appr}. Numerically, one can check that \Eq{eq:PDF:stocha:phiend:appr} is slightly better.

Fifth, the PDF of coarse-grained curvature perturbations decays exponentially as $\ee^{-\zeta_{\mathrm{cg}}}$, \ie much slower than the Gaussian decay $\ee^{-\zeta_{\mathrm{cg}}^2}$. Since PBHs form along the tail of these distributions, we expect their mass fraction to be greatly affected by this highly non-Gaussian behaviour. More precisely, on the tail, one has 
\bea
\label{eq:PDF:stoch:tail}
P(\zeta_{\mathrm{cg}},\phi)\propto \ee^{-\frac{\pi^2}{4\mu^2}\zeta_{\mathrm{cg}}}\, ,
\eea
which is given by the dominant mode $n=0$ in the expansion~(\ref{eq:stocha:HeatMethod:PDF:expansion}). Interestingly, the decay rate of the distribution is independent of $\phi$. Let us also note that another case where the PDF decays exponentially is in presence of large local non-Gaussianities, when the PDF is a $\chi^2$ distribution~\cite{Young:2013oia, Young:2015cyn}.
\section{Primordial black holes}
\label{sec:PBH}
The formalism developed so far allows one to derive the PDF of coarse-grained curvature perturbations produced during a phase of single-field slow-roll inflation. Let us now apply this result to the calculation of the mass fraction of PBHs discussed in \Sec{sec:Intro}.
\subsection{Classical limit}
\label{sec:PBH:classical}
In the classical limit detailed in \Sec{sec:ClassicalLimit}, the PDF is approximately Gaussian, see \Eq{eq:PDF:classical:NLO}, so that the considerations presented in the introduction apply. Plugging \Eq{eq:PDF:classical:NLO} into \Eq{eq:beta:def}, one has $\beta = \erfc[\zeta_\uc/(2\sqrt{v\gamma_1})]$, which is consistent with \Eq{eq:beta:erfc} as noted below \Eq{eq:PDF:classical:NLO}. In the $\beta\ll 1$ limit, this leads to 
\bea 
\label{eq:constraint:classical}
v \gamma_1 \simeq -\frac{\zeta_\uc^2}{4 \ln \beta}\, ,
\eea
where from now on, the order at which the $\gamma_i$ parameters are calculated is omitted for simplicity. Approximating $\gamma_1$ given in \Eq{eq:gamma1:nlo:def} by $\gamma_1\simeq (v/v')^3 \Delta\phi/\Mp^4$, where $\Delta\phi = \vert \phi-\phiend\vert$ is the field excursion, one obtains
\bea
\label{eq:Vconstraint:standard}
\left\vert  \frac{\Delta \phi v^4}{ {v^{\prime}}^3 \Mp^4} \right\vert \simeq - \frac{\zeta_\uc^2}{4\ln\beta(M)}\, .
\eea
In this expression, let us recall that the left-hand side must be evaluated at a value $\phi$ which is related to the PBH mass $M$ by identifying the wavenumber that exits the Hubble radius during inflation at the time when the inflaton field equals $\phi$, with the one that re-enters the Hubble radius during the radiation-dominated era when the mass contained in a Hubble patch equals $M$. For instance, with $\zeta_\uc=1$, the bound $\beta<10^{-22}$ leads to the requirement that the left-hand side of \Eq{eq:Vconstraint:standard} be smaller than $0.005$, which constrains the inflationary potential.

In passing, let us see how the first non-Gaussian correction derived in \Sec{sec:classical:nnlo} affects this result. Plugging \Eq{eq:PDF:NNLO} into \Eq{eq:beta:def}, one obtains 
\bea
\beta(M) = \erfc\left(\frac{\zeta_\uc}{2\sqrt{v\gamma_1}}\right) + \frac{\gamma_2}{4\sqrt{v\pi\gamma_1^5}}\ee^{-\frac{\zeta_\uc^2}{4 v \gamma_1}}\left(\zeta_\uc^2-2v\gamma_1\right)\, .
\eea
In the $\beta\ll 1$ limit, \ie in the $\zeta_\uc^2 \gg v \gamma_1$ limit, this reads $\beta\simeq 2 \ee^{-\zeta_\uc^2/(4 v \gamma_1)}\sqrt{v\gamma_1/\pi}/\zeta_\uc[1+\gamma_2 \zeta_\uc^3/(8 v \gamma_1^{3})]$. In this regime, one can see that the non-Gaussian correction is in fact larger than the Gaussian leading order, which signals that the non-Gaussian expansion breaks down on the far tail of the distribution. This also suggests that non-Gaussianities cannot be simply treated at the perturbative level when it comes to PBH mass fractions~\cite{Young:2015cyn}.
\subsection{Stochastic limit}
\label{sec:PBH:stochastic}
\begin{figure}[t]
\begin{center}
\includegraphics[width=0.496\textwidth]{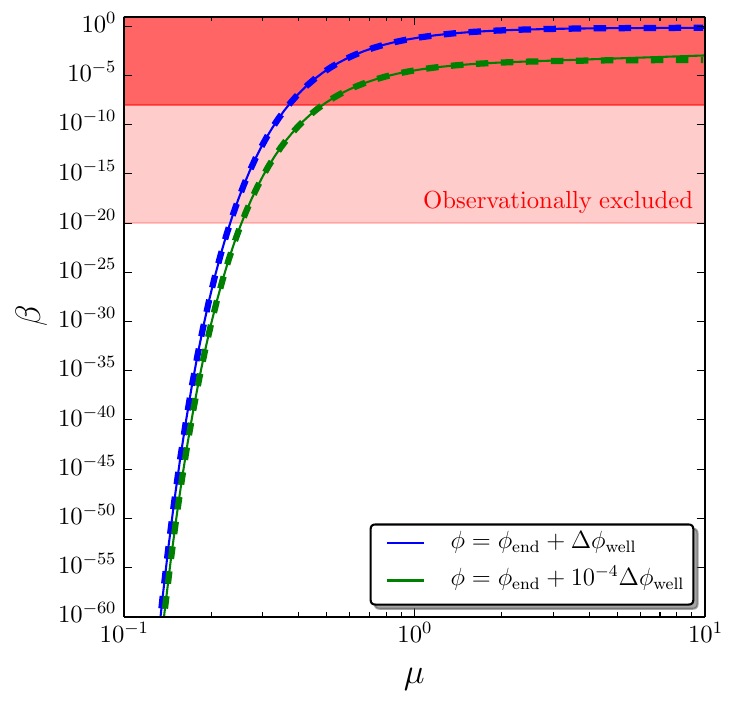}
\includegraphics[width=0.496\textwidth]{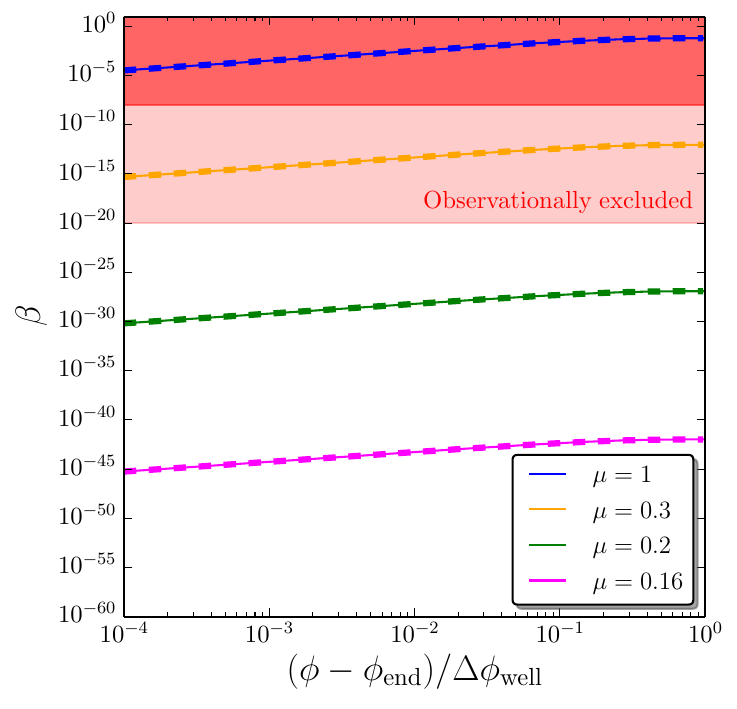}	
\caption{Mass fraction $\beta$ of primordial black holes in the quantum diffusion dominated regime. The left panel displays $\beta$ evaluated at $\phi=\phi_\uend+\Delta\phiwell$ (blue), \ie at the reflective boundary of the quantum well, and at $\phi=\phi_\uend+10^{-4}\Delta\phiwell$, \ie close to the absorbing boundary of the quantum well, as a function of $\mu=\Delta\phiwell/(\sqrt{v_0}\Mp)$. In the right panel, $\beta$ is plotted as a function of $\phi$ for a few values of $\mu$. One can see that the mass fraction depends very weakly on $\phi$ but very strongly on $\mu$. In both panels, we have taken $\zeta_{\uc} = 1$, the solid lines correspond to the full expression~(\ref{eq:beta:full}) and the dashed line to the approximation~(\ref{eq:beta:stocha:appr}). The shaded region is excluded by observations, the light shaded area roughly corresponds to constraints for PBH masses between $10^{9}\mathrm{g}$ and $10^{16}\mathrm{g}$, the dark shaded area for PBH masses between $10^{16}\mathrm{g}$ and $10^{50}\mathrm{g}$ (see discussion in \Sec{sec:Intro}).} 
\label{fig:beta}
\end{center}
\end{figure}
Let us now see how the constraint~(\ref{eq:Vconstraint:standard}) changes in the presence of large quantum diffusion, as considered in \Sec{sec:StochasticLimit}. In this case, the PDF of coarse-grained curvature perturbations $\zeta_{\mathrm{cg}} = \delta N_{\mathrm{cg}} = \N-\langle \N \rangle$ can be obtained from \Eq{eq:stocha:HeatMethod:PDF:expansion},
\bea
\label{eq:beta:full}
\beta(M) = \frac{4}{\pi} \sum^{\infty}_{n=0} \frac{1}{\left( n+\frac{1}{2} \right)}\sin{\left[ \pi \left( n + \frac{1}{2} \right)x \right]} \exp{\left\lbrace -\pi^2 \left( n + \frac{1}{2}\right)^2 \left[ x \left( 1 - \frac{x}{2} \right) + \frac{\zeta_{c}}{\mu^2}\right] \right\rbrace} \, .
\eea 
In this expression, we have replaced $\langle \N \rangle = f_1 = \mu^2 x(1-x/2)$ which can be obtained by setting the potential to a constant in \Eq{eq:fn:generalsolution}. Let us recall that $x=(\phi-\phiend)/\Delta\phiwell$ and that $M$ and $\phi$ are related as explained below \Eq{eq:Vconstraint:standard}. When $x=0$, \ie when $\phi=\phi_\uend$, \Eq{eq:beta:full} yields $\beta=0$, which is consistent with the fact that the PDF of $\zeta_{\mathrm{cg}}$ is a Dirac distribution in this case.

The mass fraction~(\ref{eq:beta:full}) depends only on $\phi$, $\mu$ and $\zeta_\uc$. It is displayed in \Fig{fig:beta} for $\zeta_\uc=1$, as a function of $\mu$ for $x=1$, \ie $\phi=\phi_\uend+\Delta\phiwell$, and $x=10^{-4}$, \ie $\phi=\phi_\uend+10^{-4}\Delta\phiwell$, in the left panel, and as a function of $\phi$ for a few values of $\mu$ in the right panel. One can see that $\beta$ depends only weakly on $\phi$ but very strongly on $\mu$, which is constrained to be at most of order one. More precisely, if one assumes that $\zeta_\uc\gg \mu^2$ so that $\zeta_\uc $ is well within the tail of the distribution and one can keep only the mode $n=0$ in \Eq{eq:beta:full}, as was done when deriving \Eq{eq:PDF:stoch:tail}, one has
\bea
\label{eq:beta:stocha:appr}
\beta(M) \simeq \frac{8}{\pi}\sin\left(\frac{\pi x}{2}\right)\ee^{-\frac{\pi^2}{8}\left[x(2-x)\right]+\frac{2\zeta_\uc}{\mu^2}}\, .
\eea
This expression is superimposed to the full result~(\ref{eq:beta:full}) in \Fig{fig:beta} where one can see that it provides a very good approximation even when the condition $\zeta_\uc\gg \mu^2$ is not satisfied. This is because, in \Eq{eq:beta:full}, higher terms in the sum are not only suppressed by higher powers of $\ee^{-\zeta_\uc^2/\mu^2}$ but also by higher powers of $\ee^{-\pi^2x(1-x/2)}$, so that \Eq{eq:beta:stocha:appr} is an excellent proxy for all values of $\mu$ except if $x$ is tiny. With $x=1$, it gives rise to
\bea
\label{eq:stocha:constraint:mu}
\mu^2 = \frac{\Delta\phi_{\mathrm{well}}^2}{v_0\Mp^2} = -\frac{2\zeta_\uc}{1+\frac{8}{\pi^2}\ln\left(\frac{\pi}{8}\beta\right)}\, .
\eea

Several comments are in order regarding this result. First, with $\zeta_\uc=1$, $\beta<10^{-24}$ gives rise to $\mu<0.21$ and $\beta<10^{-5}$ gives rise to $\mu<0.47$. The requirement that $\mu$ be smaller than one is therefore very generic and rather independent of the level of the constraint on $\beta$ or the precise value chosen for $\zeta_\uc$. Since $v_0$ needs to be smaller than $10^{-10}$ to satisfy the upper bound~\cite{Ade:2015lrj} on the tensor-to-scalar ratio in the CMB observational window, this also means that $\Delta\phiwell$ cannot exceed $\sim 10^{-5}\Mp$.

Second, \Eq{eq:stocha:constraint:mu} should be compared with its classical equivalent, \Eq{eq:Vconstraint:standard}. In the left-hand sides of these formulae, the scalings with $\Delta\phi$ and $v$ are not the same. In particular, while the PBH mass fraction increases with the energy scale $v$ in the classical picture, in the stochastic limit, it goes in the opposite direction. One should also note that when the potential is exactly flat, $v'=0$, the classical result diverges, but the stochastic one remains finite. In the right-hand sides, the scaling with $\zeta_\uc$ is also different, since the shape of the PDF $P(\zeta_{\mathrm{cg}})$ is not the same (it has a Gaussian decay in the classical case and an exponential decay in the stochastic one). The expressions~(\ref{eq:Vconstraint:standard}) and~(\ref{eq:stocha:constraint:mu}) are therefore very different, and thus translate into very different constraints on the inflationary potential. 

Third, as mentioned below \Eq{eq:beta:full}, the mean number of \efolds~realised across the quantum well is of order $\mu^2$, 
\bea
\langle \N \rangle = \mu^2 x \left(1-\frac{x}{2}\right)\, .
\eea
The conclusion one reaches is therefore remarkably simple: either the region dominated by stochastic effects is much less than one \efold~long and PBHs are not overproduced ($\mu\ll 1$), or it is much more than one \efold~long and PBHs are overproduced ($\mu\gg 1$). Interestingly, heuristic arguments lead to a similar conclusion in \Ref{GarciaBellido:1996qt}, in the context of hybrid inflation.

Fourth, in terms of the power spectrum, since \Eq{eq:Pzeta:stochaDeltaN} gives $\calP_\zeta = f_2'/f_1'- 2f_1$, with $f_1$ given above and $f_2 = 2 \mu^4 x(1 - x^2/2 + x^3/8)/3$ as can be obtained by setting the potential to a constant in \Eq{eq:fn:generalsolution}, one has 
\bea
\label{eq:Pzeta:stochasticLimit}
\calP_\zeta = \frac{2\mu^2}{3}\left(1-x\right)^2\, ,
\eea
so $\mu^2$ is also the amplitude of the power spectrum. With $\beta<10^{-22}$, the constraint~(\ref{eq:stocha:constraint:mu}) on $\mu$ translates into $\calP_\zeta<1.6\times 10^{-2}$ for the value of the power spectrum close to the end of inflation. However, contrary to the classical condition $\calP_\zeta \Delta N<10^{-2}$ recalled below \Eq{eq:Pzetaconstraint:standard}, this constraint does not involve the number of \efolds~since here, a single parameter, $\mu$, determines everything: the mean number of \efolds, the power spectrum amplitude, and the mass fraction. 
\subsection{Recipe for analysing a generic potential}
So far, we have calculated the PBH mass fraction produced in the classical limit and when the inflaton field dynamics are dominated by quantum diffusion. In order to analyse a generic potential, it remains to determine where both limits apply. This can be done by comparing the NLO and NNLO results in the classical limit to estimate the conditions under which the classical expansion is under control. For instance, comparing \Eqs{eq:gamma1:nlo:def} and~(\ref{eq:gamma:nnlo:def}) for $\gamma_1$, which gives the mass fraction $\beta$ at NLO as explained in \Sec{sec:PBH:classical}, one can see that $\vert \gamma_1^\nlo - \gamma_1^\nnlo \vert \ll \gamma_1^\nlo $ if $v\ll 1$ and $\vert v^2 v''/{v'}^2 \vert \ll 1$. The first condition is always satisfied, since as already pointed out, $v$ needs to be smaller than $10^{-10}$ to satisfy the upper bound~\cite{Ade:2015lrj} on the tensor-to-scalar ratio in the CMB observational window. The second condition defines our ``classicality criterion''~\cite{Vennin:2015hra}
\bea
\label{eq:eta_class}
\eta_{\mathrm{class}} \equiv \left \vert \frac{v^2 v^{''}}{{v'}^2} \right\vert\, .
\eea
When $\eta_{\mathrm{class}}\ll 1$, the classical expansion is under control, at least at NNLO, and one can use the results of  \Sec{sec:PBH:classical}. Of course, the classical expansion could a priori break down at NNNLO even with $\eta_{\mathrm{class}}\ll 1$, but since higher-order corrections are suppressed by higher powers of $v$, such a situation is in practice very contrived, and $\eta_{\mathrm{class}}$ provides a rather generic criterion. When $\eta_{\mathrm{class}}\gg 1$, one is far from the classical regime, quantum diffusion dominates the inflaton field dynamics and the results of \Sec{sec:PBH:stochastic} apply. When $\eta_{\mathrm{class}}$ is of order one, a full numerical treatment is required. The ``recipe'' for analysing a generic potential is therefore the following:
\begin{itemize}
\item calculate $\eta_{\mathrm{class}}$ given by \Eq{eq:eta_class} and identify the regions of the potential where $\eta_{\mathrm{class}}\ll 1$ and  $\eta_{\mathrm{class}}\gg 1$;
\item in the regions where $\eta_{\mathrm{class}}\ll 1$, make use of the constraint from \Eq{eq:constraint:classical};
\item in the ``quantum wells'' defined by $\eta_{\mathrm{class}}\gg 1$, make use of the constraint from \Eq{eq:stocha:constraint:mu}.
\end{itemize}

In the following, we illustrate this calculational programme with two examples and check its validity .
\subsection{Example 1: $V\propto 1+\phi^p$}
\label{sec:example:1_plus_phi_to_the_p}
We first consider the case where PBHs can form at scales that exit the Hubble radius towards the end of inflation, where the potential can be approximated by a Taylor expansion around $\phi=0$ where inflation is assumed to end ($\phi_\uend=0$), so
\bea
\label{eq:pot:expansEnd}
v=v_0\left[1+\left(\frac{\phi}{\phi_0}\right)^p\right]\, .
\eea
In this model, inflation does not end by slow-roll violation but another mechanism must be invoked~\cite{Linde:1991km, Linde:1993cn, Copeland:1994vg, Renaux-Petel:2015mga, Renaux-Petel:2017dia}. We also assume that the potential is in the vacuum-dominated regime for the range of field values relevant for PBH formation, so that $\phi\ll \phi_0$. A comprehensive study of this potential is performed in \App{appendix:cases} where all cases of interest are systematically identified and investigated. Here, we simply check that the calculational programme sketched above allows us to recover the main results.

In order to describe the model~(\ref{eq:pot:expansEnd}) in terms of the situation depicted in \Fig{fig:sketch2}, one has to assess $\Delta\phiwell$, which marks the boundary between the classical and the stochastic regimes. In the vacuum-dominated approximation, \Eq{eq:eta_class} gives rise to $\eta_{\mathrm{class}} \simeq (p-1) v_0 (\phi/\phi_0)^{-p}/p$, which is of order one when $\phi = \Delta\phiwell$ with
\bea
\label{eq:phiwell:expansEnd}
\Delta\phiwell \simeq  \phi_0 v_0^{\frac{1}{p}}\, .
\eea
Since $(\Delta\phiwell/\phi_0)^p=v_0\ll 1$, the vacuum-dominated condition is always satisfied at this transition point. However, the slow-roll conditions are not always met, and in \App{appendix:cases} it is shown that slow roll is indeed violated at $\phi = \Delta\phiwell$ if $\phi_0/\Mp < v_0^{(p-2)/(2p)}$, unless $p=1$ for which slow-roll is violated if $\phi_0< \Mp$. In such cases, the expansion~(\ref{eq:pot:expansEnd}) fails to cover the whole quantum well and higher-order terms in the potential must be included for a consistent analysis. Otherwise, we can keep following the recipe given above.

In the classical regime, $\phi\gg \Delta\phiwell$, \Eq{eq:constraint:classical} applies, where $v \gamma_1$ is given by \Eq{eq:gamma1:nlo:def}. In the vacuum-dominated approximation, it reads $v\gamma_1 \simeq v_0 (\phi_0/\Mp)^4/(4p^3-3p^4)[(\phi/\phi_0)^{4-3p}-(\phi_\uend/\phi_0)^{4-3p}]$. Neglecting the contribution from $\phi_\uend$, which lies outside the validity range of the classical formula anyway, one can evaluate this expression at $\phi = \Delta\phiwell$ where the power spectrum is maximal, and combining this with \Eq{eq:constraint:classical} leads to
\bea
\label{eq:expansEnd:classConstraint}
\frac{v_0^{\frac{2}{p}-1}}{\sqrt{\vert 4p^3-3p^4 \vert }}\left(\frac{\phi_0}{\Mp}\right)^2 \simeq  \frac{\zeta_\uc}{2\sqrt{\vert \ln \beta \vert}}\, .
\eea
In the stochastic regime, combining \Eqs{eq:stocha:constraint:mu} and~(\ref{eq:phiwell:expansEnd}), one has
\bea
\label{eq:expansEnd:stochConstraint}
v_0^{\frac{2}{p}-1}\left(\frac{\phi_0}{\Mp}\right)^2 \simeq  \frac{2\zeta_\uc}{\left\vert 1+\frac{8}{\pi^2}\ln\left(\frac{\pi}{8}\beta\right)\right\vert}\, .
\eea
It is interesting to notice that up to an overall factor of order one, the two constraints~(\ref{eq:expansEnd:classConstraint}) and~(\ref{eq:expansEnd:stochConstraint}) are very similar, even though they are obtained in very different regimes that yield very different PDFs for the curvature perturbations. 

It is also important to note that the slow-roll conditions given above imply that $v_0^{2/p-1}(\phi_0/\Mp)^2\gg 1$ except if $p=1$. Therefore, if $p$ is different from $1$, either PBHs are too abundant and the model is ruled out, or slow roll is strongly violated before one exits the classical regime and one needs to go beyond the present formalism to calculate PBH mass fractions. The case $p=1$ is subtle, since \Eq{eq:eta_class} gives $\eta_{\mathrm{class}}=0$. One would thus have to extend the classical expansion of \Sec{sec:ClassicalLimit} to next-to-next-to-next to leading order (NNNLO) to determine what the first stochastic correction is and under which condition the classical approximation holds, and investigate numerically the regime where it does not. This goes beyond the scope of the present paper and we leave this study for future work.

\begin{figure}[t]
\begin{center}
\includegraphics[width=0.6\textwidth]{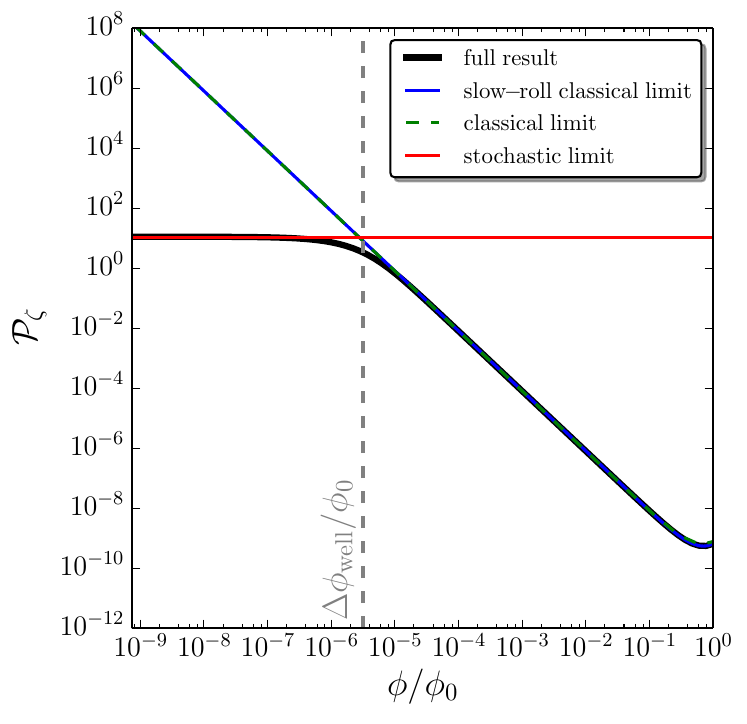}
\caption{Power spectrum of curvature perturbations $\calP_\zeta$ produced in the potential~(\ref{eq:pot:expansEnd}) with  $p=2$, $v_0 = 10^{-11}$, $\phi_0=4 \Mp$ and $\phiuv=10^{4}\phi_0$ (solid black line). The blue line corresponds to the slow-roll classical limit~(\ref{eq:classicalPS}), while the green dashed line is obtained from solving the full Klein-Gordon equation. The red line corresponds to the stochastic limit assuming the potential is exactly flat for $\phi<\Delta\phiwell$ and that a reflective wall is located at $\phi = \Delta\phiwell$.  The value of $\Delta\phiwell$ obtained from requiring $\eta_{\mathrm{class}}=1$ is displayed with the grey vertical dotted line and delimitates the classical and stochastic regimes.} 
\label{fig:quad:power:spectrum}
\end{center}
\end{figure}
In passing, let us check that approximating the full potential~(\ref{eq:pot:expansEnd}) as a piecewise function consisting of a constant piece and a classical one, separated at $\phi=\Delta\phiwell$, is numerically justified. In \Fig{fig:quad:power:spectrum}, we show the power spectrum computed numerically from \Eqs{eq:fn:generalsolution} and~(\ref{eq:Pzeta:stochaDeltaN}), which gives $\calP_\zeta = f_2'/f_1'- 2f_1$, in the potential~(\ref{eq:pot:expansEnd}) with $p=2$, $v_0 = 10^{-11}$, $\phi_0=4 \Mp$ and $\phiuv=10^{4}\phi_0$ (solid black line). The blue line corresponds to the slow-roll classical limit~(\ref{eq:classicalPS}), and the green dashed line is obtained from solving the full Klein-Gordon equation. The agreement of this solution with the slow-roll formula confirms that the slow-roll conditions are satisfied for the parameters used in this example. The red line corresponds to the stochastic limit~(\ref{eq:Pzeta:stochasticLimit}) $\calP_\zeta = 2\mu^2/3$ at $\phi=0$, where $\mu$ is given by \Eqs{eq:def:mu} and~(\ref{eq:phiwell:expansEnd}), which yields $\calP_\zeta \sim 2 (\phi_0/\Mp)^2 v_0^{2/p-1}/3$. One can see that both limits are correctly reproduced, and that the value of $\Delta\phiwell$ obtained in \Eq{eq:phiwell:expansEnd} from our classicality criterion $\eta_{\mathrm{class}}<1$, and displayed with the grey vertical dotted line, indeed separates the two regimes. In \App{appendix:cases}, an analytical expression for $\calP_\zeta$ in the regime $\phi\ll\Delta\phiwell$ is derived, and one finds $\calP_\zeta = 2\Gamma^2(1/p)v_0^{2/p-1}(\phi_0/\Mp)^2/p^2$, see \Eq{stochasticpower}, where $\Gamma$ is the gamma function. Up to an overall numerical constant of order one, one recovers the result obtained from simply assuming the potential to be exactly flat until $\phi=\Delta\phiwell$, where $\eta_{\mathrm{class}}=1$, and setting a reflective wall there. This confirms the validity of this approach.
\subsection{Example 2: running-mass inflation}
\begin{figure}[t]
\begin{center}
\includegraphics[width=0.6\textwidth]{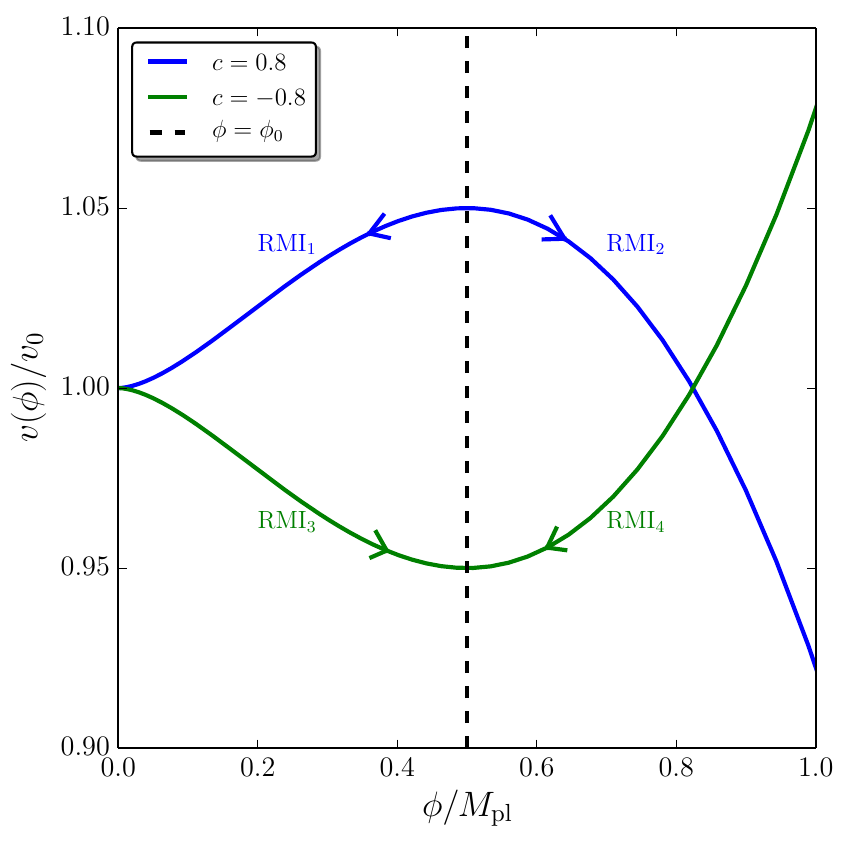}		
\caption{The potential~(\ref{eq:RMI:potential}) for running-mass inflation (RMI) with $\phi_{0} = 0.5\Mp$. The blue curve takes $c = 0.8$ and the green curve takes $c = -0.8$ (these values may not be physical but they have been chosen to produce a clear plot). RMI is shown to have four possible realisations (RMI${}_1$, RMI${}_2$, RMI${}_3$ and RMI${}_4$), depending on the sign of $c$ and on whether $\phi$ is initially smaller or larger than $\phi_0$. Except for RMI${}_2$, the potential flattens as inflation proceeds, which can lead to the formation of PBHs for scales exiting the Hubble radius towards the end of inflation.} 
\label{fig:RMI:Potential}
\end{center}
\end{figure}
Let us now consider another example, running-mass inflation (RMI) \cite{Stewart:1996ey}, where the inflationary potential is given by
\bea
\label{eq:RMI:potential}
v\left( \phi \right) = v_{0} \left\lbrace 1 - \frac{c}{2}\left[ -\frac{1}{2} + \ln{\left( \frac{\phi}{\phi_{0}}\right)} \right] \frac{\phi^2}{\Mp^2}\right\rbrace\, .
\eea
In this expression, $c$ is a dimensionless coupling constant assumed to be much smaller than one, $c\ll 1$ (more precisely, as discussed in \Ref{Martin:2013tda}, within supersymmetry, natural values of $c$ are $c \simeq 10^{-2}$ to $10^{-1}$ for soft masses values matching the energy scale of inflation), and $\phi_0$ must be sub-Planckian, $\phi_0\ll \Mp$. 

The potential~(\ref{eq:RMI:potential}) is displayed in \Fig{fig:RMI:Potential}, where one can see that depending on the sign of $c$ and on whether $\phi$ is initially smaller or larger than $\phi_0$, inflation can proceed in four regimes~\cite{Covi:1998mb}, that we denote RMI${}_1$, RMI${}_2$, RMI${}_3$ and RMI${}_4$. In RMI${}_1$, $c > 0$, $\phi < \phi_{0}$ and $\phi$ decreases towards $\phi = 0$ as inflation proceeds. RMI${}_2$ also has $c > 0$ but in this case $\phi > \phi_{0}$ and $\phi$ increases away from $\phi_{0}$ throughout inflation. RMI${}_3$ and RMI${}_4$ both have $c < 0$, but RMI${}_3$ has $\phi < \phi_{0}$ with $\phi$ increasing towards $\phi_0$ during inflation, while RMI${}_4$ has $\phi > \phi_{0}$ and $\phi$ decreases towards $\phi_0$ during inflation. In RMI${}_1$, RMI${}_3$ and RMI${}_4$, the potential flattens as inflation proceeds, which may lead to the production of PBHs at scales that exit the Hubble radius towards the end of inflation, as studied in \Refs{Leach:2000ea, Drees:2011hb, Akrami:2016vrq}. 
The width of the ``quantum well'' in these cases is determined by the condition $\eta_{\mathrm{class}}>1$, where $\eta_{\mathrm{class}}$ is given by \Eq{eq:eta_class}. In the vacuum-dominated regime, it reads
\bea
\eta_{\mathrm{class}} \simeq \frac{v_0}{\vert c \vert}\frac{\Mp^2}{\phi^2}\frac{\left\vert 1+\ln\left(\frac{\phi}{\phi_0}\right)\right\vert}{\ln^2\left(\frac{\phi}{\phi_0}\right)}
\, .
\eea

For RMI${}_1$, the equation $\eta_{\mathrm{class}}(\phiwell)=1$ yields $\phiwell/\phi_0 \sqrt{\vert \ln(\phiwell/\phi_0 ) \vert} = \sqrt{v_0/c} \Mp/\phi_0$, where we have assumed that $\phiwell \ll \phi_0$ so that $\vert \ln(\phiwell/\phi_0)\vert \gg 1$. This can be solved as
\bea
\label{eq:phiwell:RMI1}
\phiwell = \phi_0 \exp\left[\frac{1}{2 }W_{-1}\left( -2 \frac{v_0}{ c  }\frac{\Mp^2}{\phi_0^2}\right)\right]\, ,
\eea
where $W_{-1}$ is the $-1$ branch of the Lambert function~\cite{Olver:2010:NHM:1830479:lambert}. The approximation $\vert \ln(\phiwell/\phi_0)\vert \gg 1$ is satisfied when the argument of the Lambert function in \Eq{eq:phiwell:RMI1} is much smaller than one, which is typically the case for the values of $v_0$, $c$ and $\phi_0$ considered in the literature~\cite{Martin:2013tda, Martin:2013nzq}. In this limit, one can Taylor expand the Lambert function according to $W_{-1}(-x)\simeq \ln x$ when $x\ll 1$, which gives rise to $\phiwell \simeq \sqrt{2v_0/c} \Mp$, and hence
\bea
\label{eq:Deltaphiwell:RMI1}
\Delta\phiwell = \left\vert \phiwell \right\vert \simeq   \sqrt{\frac{2 v_0}{c}}\Mp\, .
\eea
In this expression, we have assumed that inflation terminates at $\phi=0$ (otherwise, $\phiend$ must be subtracted from the right-hand side). Making use of \Eq{eq:def:mu}, this leads to
\bea
\label{eq:MU:RMI1}
\mu^2\simeq \frac{2}{c} \gg 1\, .
\eea
The result is remarkably simple since it depends only on the coupling constant $c$, and on neither $v_0$ nor $\phi_0$. As explained at the beginning of this section, $c$ is always much smaller than one, which implies that $\mu\gg 1$ and according to the discussion of \Sec{sec:PBH:stochastic}, PBHs are too abundant in this case. One concludes that the stochastic regime of the potential cannot be explored in RMI${}_1$, \ie $\phiend$ should be at least of the order of $\Delta\phiwell$ given in \Eq{eq:Deltaphiwell:RMI1}.

For RMI${}_2$, the potential does not flatten as inflation proceeds so the inflaton field dynamics is never dominated by quantum diffusion for scales smaller than those probed in the CMB.

For RMI${}_3$ and RMI${}_4$, assuming that $\phiwell$ is very close to $\phi_0$ so that $\vert \ln(\phiwell/\phi_0)\vert \ll 1$, the equation $\eta_{\mathrm{class}}(\phiwell)=1$ reduces to $\phiwell/\phi_0 \vert \ln(\phiwell/\phi_0 ) \vert = \sqrt{v_0/\vert c \vert} \Mp/\phi_0$. This can be solved as
\bea
\label{eq:phiwell:RMI34}
\phiwell = \phi_0 \exp\left[W_0\left(\mp \sqrt{\frac{v_0}{\vert c \vert}}\frac{\Mp}{\phi_0}\right)\right]\, ,
\eea
where $W_0$ is the principal branch of the Lambert function, and its argument comes with a minus sign in RMI${}_3$ and with a plus sign in RMI${}_4$. The approximation $\vert \ln(\phiwell/\phi_0)\vert \ll 1$ is satisfied when the argument of the Lambert function in \Eq{eq:phiwell:RMI34} is much smaller than one, which is the same condition as the one coming from $\phiwell\ll\phi_0$ in RMI${}_1$. In this limit, one can Taylor expand the Lambert function as $W_0(x)\simeq x$ when $x\ll 1$. One obtains $\phiwell \simeq \phi_0\mp \Mp\sqrt{v_0/\vert c \vert}$, and hence
\bea
\label{eq:Deltaphiwell:RMI34}
\Delta\phiwell = \left\vert \phiwell-\phi_0 \right\vert \simeq  \Mp\sqrt{\frac{v_0}{\left\vert c \right\vert}}\, .
\eea
Up to a factor $\sqrt{2}$, this expression is the same as \Eq{eq:Deltaphiwell:RMI1}. This leads to
\bea
\label{eq:MU:RMI34}
\mu^2\simeq \frac{1}{\left\vert c \right\vert} \gg 1\, ,
\eea
and the same conclusions as the ones drawn for RMI${}_1$ apply, namely that one cannot explore the quantum well of the potential without producing too many PBHs, so $\vert \phiend-\phi_0\vert $ should be at least of order $\Delta\phiwell$ given in \Eq{eq:Deltaphiwell:RMI34}.

If this is indeed the case, the classical approximation is valid throughout the entire period of inflation, and \Eq{eq:classicalPS} gives rise to
\bea
\calP_\zeta \simeq 2\frac{v_0}{c^2}\frac{\Mp^2}{\phi^2}\ln^{-2}\left(\frac{\phi}{\phi_0}\right)\, .
\eea
When this expression is evaluated at $\phiwell$, given by \Eq{eq:phiwell:RMI1} for RMI${}_1$ and by \Eq{eq:phiwell:RMI34} for RMI${}_3$ and RMI${}_4$, one finds
\bea
\calP_\zeta\left(\phiwell\right) \simeq
\begin{cases}
\frac{4}{c} &\mathrm{in}\ \mathrm{RMI}_1\\
\frac{2}{\left\vert c\right\vert } &\mathrm{in}\ \mathrm{RMI}_3\ \mathrm{and}\ \mathrm{RMI}_4
\end{cases}\, .
\eea
It is interesting to notice that, up to an overall numerical constant of order one, this also corresponds to the stochastic limit~(\ref{eq:Pzeta:stochasticLimit}), $\calP_\zeta \sim \mu^2 \sim 1/c$, where one makes use of \Eqs{eq:MU:RMI1} and~(\ref{eq:MU:RMI34}). This is similar to what was found in \Sec{sec:example:1_plus_phi_to_the_p}. Since $\vert c \vert \ll 1$, this means that the classical power spectrum is already larger than one when one enters the quantum well. This implies that, for this model, analyses relying on the classical formalism only should exclude the quantum well (even if not for the correct reason) and should therefore be valid. Let us however stress that the approach developed in this work was necessary in order to check the consistency of the standard results.
\section{Conclusion}
\label{sec:Conclusion}
Let us now summarise our main findings. Making use of the stochastic-$\delta N$ formalism, we have developed a calculational framework in which the PDF of the coarse-grained curvature perturbations produced during inflation can be derived exactly, even in the presence of large quantum backreaction on the inflaton field dynamics.  More precisely, we have proposed two complementary methods, one based on solving an ordinary differential equation for the characteristic function of the PDF, and the other based on solving a heat equation for the PDF directly. We have shown that depending on the problem one considers, the method to be preferred can vary. We have then derived a classicality criterion that determines whether the effects of quantum diffusion are important or not. When this is not the case, \ie in the classical limit, we have developed an expansion scheme that not only recovers the standard Gaussian PDF at leading order, but also allows one to calculate the first non-Gaussian corrections to the usual result. In the opposite limit, \ie when quantum diffusion plays the dominant role in the field dynamics, we have found that the PDF follows an elliptic theta function, whose tail decays only exponentially, and which is fully characterised by a single parameter, given by $\mu^2 = \Delta\phiwell^2/(v_0\Mp^2)$. This parameter measures the ratio between the squared width of the quantum well and its height, in Planck mass units. The mean number of \efolds~realised across the quantum well, the amplitude of the power spectrum, and, if $\zeta_\uc \sim 1$, the inverse log of the PBH mass fraction, are all of order $\mu^2$. Therefore, observational constraints on the abundance of PBHs put an upper bound on $\mu^2$ that is of order one, and imposes that one cannot spend more than $\sim 1$ \efold~in regions of the potential dominated by quantum diffusion. For a given potential, one must therefore determine whether a diffusion dominated quantum well exists, and check that its width squared is smaller than its height. Finally, we have illustrated our calculational programme with two examples.

We now mention a few of the new and interesting research directions that open up as a consequence of the results obtained in this work:
\begin{itemize}[leftmargin=*]
\item First, we have shown that the effects of quantum diffusion on the PDF of curvature perturbations and on the mass fraction of PBHs can be dramatic in regions of the potential where the classical approximation breaks down. This implies that some of the constraints on inflationary models derived in the literature, from non-observations of PBHs and using only the classical approximation, may have to be revised. This could have important consequences for these models.

\item Second, we have seen that even when the classical approximation is under control close to the maximum of the PDF, it fails on its tail, where deviations from Gaussianity cannot be simply described by increasing the order at which the classical expansion is performed. Since PBHs are precisely sourced by the tail of the distribution of curvature perturbations, this implies that a more thorough investigation of quantum diffusion effects on PBH mass fraction may be required, even in what is referred to as the ``classical'' regime in the present work.

\item Third, there are cases where slow roll is violated when scales smaller than those probed in the CMB exit the Hubble radius and the standard PBH calculation does not apply, as recently pointed out in \Refs{Kannike:2017bxn, Germani:2017bcs, Motohashi:2017kbs}. This, for instance, happens when the inflationary potential has a flat inflection point \cite{Garcia-Bellido:2017mdw}, around which slow roll is transiently violated and one can even enter an ultra slow-roll phase. However, close to the inflection point, quantum diffusion plays an important role and this also needs to be included. This requires the formalism presented in this work to be extended beyond the slow-roll approximation~\cite{Grain:2017dqa}.

\item Fourth, it has recently been shown~\cite{Assadullahi:2016gkk, Vennin:2016wnk} that in presence of multiple fields, the effects of quantum diffusion can be even more drastic. Formation of PBHs in multi-field models of inflation, such as hybrid inflation~\cite{GarciaBellido:1996qt, Bugaev:2011qt, Bugaev:2011wy, Clesse:2015wea, Kawasaki:2015ppx}, would therefore be interesting to study with our formalism.

\item Fifth and finally, there are other astrophysical objects for which the knowledge of the full probability distribution of cosmological perturbations produced during inflation is important, such as ultra-compact mini-halos~\cite{Shandera:2012ke}. Using our results to calculate the abundance of such objects is also an interesting prospect.
\end{itemize}
\begin{acknowledgments}
It is a pleasure to thank Chris Byrnes and Sam Young for enjoyable discussions and useful comments.  V.V. acknowledges funding from the European Union's Horizon 2020 research and innovation programme under the Marie Sk\l odowska-Curie grant agreement N${}^0$ 750491. V.V., H.A. and D.W. acknowledge financial support from STFC grant ST/N000668/1.
\end{acknowledgments}
\appendix
\section{Elliptic theta functions}
\label{appendix:identities}
In \Sec{sec:StochasticLimit}, the PDF of coarse-grained curvature perturbations is expressed in terms of elliptic theta functions. In this appendix, we define these special functions and give some of their properties that are relevant for the considerations of \Sec{sec:StochasticLimit}. There are four elliptic theta functions, defined as~\cite{Olver:2010:NHM:1830479:theta, Abramovitz:1970aa:theta}
\bea
\vartheta_1\left(z,q\right) &= 2 \sum_{n=0}^\infty (-1)^n q^{\left(n+\frac{1}{2}\right)^2}\sin\left[\left(2n+1\right)z\right]\, ,\\
\vartheta_2\left(z,q\right) &= 2 \sum_{n=0}^\infty  q^{\left(n+\frac{1}{2}\right)^2}\cos\left[\left(2n+1\right)z\right]\, ,\\
\vartheta_3\left(z,q\right) &= 1+2 \sum_{n=1}^\infty q^{n^2}\cos\left(2nz\right)\, ,\\
\vartheta_4\left(z,q\right) &= 1+2 \sum_{n=1}^\infty (-1)^nq^{n^2}\cos\left(2nz\right)\, .
\eea
By convention, $\vartheta_i^\prime$ denotes the derivative of $\vartheta_i$ with respect to its first argument $z$. For instance, one has
\bea
\label{eq:theta1prime:def}
\vartheta_1^\prime\left(z,q\right) = 2 \sum_{n=0}^\infty (-1)^n q^{\left(n+\frac{1}{2}\right)^2}\left(2n+1\right)\cos\left[\left(2n+1\right)z\right]\, ,
\eea
which appears in \Eq{eq:PDF:phiwall:thetaElliptic}. As another example, one has
\bea
\label{eq:theta4primeprime:def}
\vartheta_4^{\prime\prime}\left(z,q\right) = -8\sum_{n=1}^\infty (-1)^nq^{n^2}n^2\cos\left(2nz\right)\, ,
\eea
which is used in \Eq{eq:PDF:stocha:phiend:appr}. As a third example, one has
\bea
\label{eq:theta2prime:def}
\vartheta_2^{\prime}\left(z,q\right) = -2 \sum_{n=0}^\infty  q^{\left(n+\frac{1}{2}\right)^2}\left(2n+1\right)\sin\left[\left(2n+1\right)z\right]\, ,
\eea
which appears in \Eq{eq:PDF:thetatwo}.  As a last example, one has
\bea
\label{eq:theta2primeprime:def}
\vartheta_2^{\prime\prime}\left(z,q\right) = -2 \sum_{n=0}^\infty  q^{\left(n+\frac{1}{2}\right)^2}\left(2n+1\right)^2\cos\left[\left(2n+1\right)z\right]\, ,
\eea
which is used in \Eq{eq:PDF:stocha:phiend:appr:2}. The function $\vartheta_i(z,q)$ is noted \texttt{EllipticTheta[i,z,q]} in Mathematica and $\vartheta_i^\prime(z,q)$ is noted \texttt{EllipticThetaPrime[i,z,q]}.

Let us now show that the different expressions for $P(\N,\phi=\phiend+\Delta\phiwell)$ obtained in \Sec{sec:StochasticLimit} in the stochastic dominated regime are equivalent. A first expression is given by \Eq{eq:PDF:phiwall:thetaElliptic}, a second expression can be derived by plugging $x=1$ into \Eq{eq:stocha:HeatMethod:PDF:expansion} and making use of \Eq{eq:theta1prime:def}, and a third expression is given by plugging $x=1$ in \Eq{eq:PDF:thetatwo}. The three formulae are equivalent if
\bea
\label{eq:theta:identity}
\left(\frac{\mu}{\sqrt{\pi\N}}\right)^3 \vartheta^\prime_1\left(0,\ee^{-\frac{\mu^2}{\N}}\right) = 
\vartheta_{1}'\left( 0, \ee^{ -\frac{\pi^2}{\mu^2} \N} \right)  = 
-\vartheta_{2}'\left( \frac{\pi}{2}, \ee^{ -\frac{\pi^2}{\mu^2} \N } \right) \, .
\eea
The first equality in \Eq{eq:theta:identity} can be shown from the Jacobi identity for a modular transformation of the first elliptic theta function, see Eq.~(20.7.30) of \Ref{NIST:DLMF},
\bea
\label{app:identity:modulartrans}
\left( -i \tau \right)^{\frac{1}{2}} \vartheta_{1} \left( z, \ee^{i\pi \tau} \right) = -i \ee^{ -\frac{z^2}{\pi \tau}} \vartheta_{1} \left( -\frac{z}{\tau}, -\ee^{-\frac{i\pi}{\tau}} \right) \, .
\eea
By taking $\tau = i/(a \pi)$ and differentiating \Eq{app:identity:modulartrans} with respect to $z$, one obtains
\bea
\label{app:id:diff}
\left( \pi a \right)^{\frac{1}{2}} \vartheta_{1}' \left( z, \ee^{-\frac{1}{a}} \right) = - \frac{2iz}{a}\ee^{az^2} \vartheta_{1} \left( - i \pi a z, \ee^{-\pi^2 a} \right) + a \pi \ee^{az^2} \vartheta_{1}' \left( - i \pi a z, \ee^{-\pi^2 a} \right) \, .
\eea
Taking $z = 0$, one recovers the first equality in \Eq{eq:theta:identity}. The second equality in \Eq{eq:theta:identity} simply follows from \Eqs{eq:theta1prime:def} and~(\ref{eq:theta2prime:def}).
\section{Detailed analysis of the model $V\propto 1+\phi^p$}
\label{appendix:cases}
\addtocontents{toc}{\protect\setcounter{tocdepth}{-1}}
In this appendix, we present a detailed analysis of the model discussed in \Sec{sec:example:1_plus_phi_to_the_p}, where the inflationary potential is of the form
\bea
\label{app:eq:def:potential}
v(\phi) = v_{0}\left[ 1 + \left( \frac{\phi}{\phi_{0}} \right)^{p} \right] \, .
\eea

In order to use the slow-roll approximation, one needs to check that the slow-roll conditions~\cite{liddle_lyth_2000}, $\Mp^{2}(v'/v)^2 \ll 1$, $\Mp^{2}|v''/v| \ll 1$, and $\Mp^{4}|v'''v'/v^{2}| \ll 1$, are satisfied. Here, we use the three first slow-roll conditions only, since these are the only ones currently constrained by observations. In order to satisfy the third condition, one requires a condition involving the position of $\phi$ with respect to
\bea
\label{constraint3}
\phi_{\text{sr1}} \equiv {\phi_{0}} \left( \frac{\phi_{0}}{\Mp}\right) ^{\frac{2}{p - 2}} \, ,
\eea
\ie $\phi \ll \phi_{\text{sr1}}$ if $p > 2$ and $\phi \gg \phi_{\text{sr1}}$ if $p<2$. The second slow-roll condition reduces to an equivalent condition. If $p=1$, there is no such condition and if $p=2$, it reduces to $\phi_0\gg \Mp$. The first condition constrains the position of $\phi$ with respect to
\bea
\label{constraint1}
\phi_{\text{sr2}} \equiv  \phi_0 \left(\frac{\phi_{0}}{\Mp}\right)^{\frac{1}{p - 1}} \, ,
\eea
namely $\phi\ll \phi_{\text{sr2}} $ if $p>1$ and $\phi\gg \phi_{\text{sr2}} $ if $p<1$ (if $p=1$, it reduces to $\phi_0\gg \Mp$).

As explained in \Sec{sec:example:1_plus_phi_to_the_p}, quantum diffusion plays an important role when $\eta_{\mathrm{class}}\gg 1$, where $\eta_{\mathrm{class}}$ is given in \Eq{eq:eta_class}, which leads to $\phi \ll \Delta\phiwell$,  where
\bea
\label{app:example:Deltaphiwell}
\Delta\phiwell = \phi_0 v_0^{\frac{1}{p}}\, ,
\eea
see \Eq{eq:phiwell:expansEnd}. Hereafter, we work in the vacuum-dominated regime, in which $\phi \ll \phi_0$ and $v \simeq v_{0}$. Making use of \Eq{eq:stocha:constraint:mu}, this gives rise to
\bea
\label{eq:app:example:mu}
\mu^2 = \left(\frac{\phi_0}{\Mp}\right)^2 v_0^{\frac{2}{p}-1}\, .
\eea

In the classical regime, \ie when $\phi\gg \Delta\phiwell $, the power spectrum is given by \Eq{eq:classicalPS}, which gives rise to
\bea
\label{clpower}
\mathcal{P}_{\zeta} |_{\text{cl}} = \frac{2v_{0}}{p^2}\left(\frac{\phi_{0}}{\Mp}\right)^{2}\left(\frac{\phi_{0}}{\phi}\right)^{2p - 2} \, .
\eea
Thus we see that the classical power spectrum is larger than unity when $\phi < \phicl$ if $p>1$, and $\phi >\phicl$ if $p<1$, where
\bea 
\label{clpower>1}
\phicl = {\phi_{0}} \left[ \frac{2v_{0}}{p^2}\left(\frac{\phi_{0}}{\Mp}\right)^2 \right]^{\frac{1}{2p - 2}}\, .
\eea
The number of \efolds~realised between $\phi$ and $\phiend$ can also be calculated in the classical regime using \Eq{eq:classicalPS}, and one obtains
\bea 
\label{clefolds}
N_{\text{end}} - N \simeq \frac{\phi_{0}^2}{p(p - 2)\Mp^{2}}\left[ \left( \frac{\phi_{0}}{\phiend}\right)^{p - 2} - \left( \frac{\phi_{0}}{\phi}\right)^{p - 2}\right]\, .
\eea
Note that this expression is singular for $p=2$, and this case is treated separately in \Sec{app:V_eq_1_plus_phi_to_the_p:p_eq_2}. Combining \Eqs{clpower>1} and~(\ref{clefolds}), one can rewrite the classical power spectrum as
\bea
\label{clPzeta}
\mathcal{P}_{\zeta} |_{\text{cl}} = \frac{2v_{0}}{p^2}\left(\frac{\phi_{0}}{\Mp}\right)^{2}\left[ \frac{p(2 - p)\Mp^{2}}{\phi_{0}^2} (N_{\text{end}} - N) + \left( \frac{\phi_{0}}{\phiend}\right)^{p - 2} \right]^{\frac{2p - 2}{p - 2}} \, .
\eea

In the stochastic regime,  \ie when $\phi\ll \Delta\phiwell $, the mean number of \efolds~can be computed from \Eq{eq:fn:generalsolution} for $f_1$. In the limit $\phiuv \rightarrow \infty$, and taking $\phi \ll \Delta\phiwell$, one obtains
\bea 
\label{stochasticefolds}
\left< \N \right> = \Gamma \left( \frac{1}{p} \right) \frac{\phi_{0}}{p \Mp^2v_{0}^{1 - \frac{1}{p}}} \left( \phi - \phiend \right) \, .
\eea
Similarly, using \Eq{eq:fn:generalsolution} for $f_2$, and the formula $\calP_\zeta = f_2'/f_1'- 2f_1$ from \Eq{eq:Pzeta:stochaDeltaN}, the power spectrum is given by
\bea 
\label{stochasticpower}
\mathcal{P}_{\zeta} (\phi) = \frac{2}{p^2}\Gamma^{2}\left( \frac{1}{p} \right) v_{0}^{\frac{2}{p} - 1} \left(\frac{\phi_{0}}{\Mp}\right)^2 \, .
\eea
\subsection{Case $p=2$}
\label{app:V_eq_1_plus_phi_to_the_p:p_eq_2}
We first consider the case of $p = 2$, where the slow-roll conditions reduce to $\phi_0 \gg \Mp$. Let us also note that \Eq{clefolds} is  singular for $p=2$, and that it should be replaced by
\bea
N_\uend - N \simeq \frac{\phi_0^2}{2\Mp^2} \ln\left(\frac{\phi}{\phi_\uend}\right)\, .
\eea
As noted in \Sec{sec:example:1_plus_phi_to_the_p}, the constant value found for the power spectrum in the stochastic limit, \Eq{stochasticpower}, corresponds (up to an order one prefactor) to the classical power spectrum~(\ref{clPzeta}) evaluated at $\phisto$. Therefore, when $\phi$ decreases, the stochastic and the classical result coincide until $\phi$ becomes smaller than $\phisto$, where the stochastic power spectrum saturates to a constant value and the classical power spectrum continues to increase. Since the slow-roll condition implies that  $\phi_{0} \gg \Mp$, the power spectrum is always larger than one in this regime. This can be also seen from $\phicl < \phisto$, as can be checked explicitly from \Eqs{app:example:Deltaphiwell} and~(\ref{clpower>1}). 

Furthermore, the number of \efolds~(\ref{stochasticefolds}) spent in the stochastic regime is of order $\mean{\N} = \sqrt{\pi}/2(\phi_0/\Mp)^2$, which is larger than unity. It is interesting to note that both the amplitude of the power spectrum and the number of \efolds~during which the stochastic regime depend only on $\phi_0$. PBHs are therefore overproduced in this case.
\subsection{Case $p > 2$}
We now consider values of $p$ such that $p > 2$. In this case, the slow-roll condition that is the strictest depends on whether $\phi_0$ is sub-Planckian or super-Planckian. More precisely, since $2/({p-2}) > 1/({p-1})$ for $p>2$, if $\phi_{0} < \Mp$, then $\phi_{\text{sr1}} < \phi_{\text{sr2}}$ and the slow-roll condition is given by $\phi \ll \phi_{\text{sr1}}$, and if $\phi_{0} > \Mp$ then $\phi_{\text{sr2}} < \phi_{\text{sr1}}$ and it is given by $\phi \ll \phi_{\text{sr2}}$.
\subsubsection{Case $p > 2$ and $\phi_{0} > \Mp$}
\label{app:example:p_gt_2_and_phi0_gt_Mp}
In this case, the slow-roll condition reads $\phi \ll \phi_{\text{sr2}}$. Making use of \Eqs{constraint1}, \eqref{app:example:Deltaphiwell} and \eqref{clpower>1}, one can show that
\bea
\phisto < \phicl < \phi_{\text{sr2}}\, .
\eea
As a consequence, when stochastic effects become important, the classical power spectrum is already larger than one and so quantum diffusion cannot ``rescue'' the model in this case. Stochastic effects do reduce the amount of power, but not soon enough to keep the amount of PBH below the observationally constrained level.
\subsubsection{Case $p > 2$ and $\phi_{0} < \Mp$}
In this case, the slow-roll condition reads $\phi \ll \phi_{\text{sr1}}$. Two sub-cases need to be distinguished.
\paragraph{Case $p>2$ and $v_{0}^{\frac{p - 2}{2p}}<\phi_{0}/\Mp<1 $}
$ $\\
In this case, \Eqs{constraint3}, \eqref{app:example:Deltaphiwell} and \eqref{clpower>1} lead to
\bea
\phisto < \phicl < \phi_{\text{sr1}}\, .
\eea
The situation is therefore very similar to the case $p>2$ and $\phi_{0} > \Mp$, and quantum diffusion does not sufficiently suppress PBH production. 
\paragraph{Case $p>2$ and $\phi_{0}/\Mp < v_{0}^{\frac{p - 2}{2p}}$}
$ $\\
In this case, \Eqs{constraint3}, \eqref{app:example:Deltaphiwell} and \eqref{clpower>1} give a reversed hierarchy, namely
\bea
\phi_{\text{sr1}} < \phicl < \phisto \,.
\eea
In the region where the slow-roll approximation applies, $\phi\ll \phi_{\text{sr1}}$, the classical power spectrum is therefore always larger than one. However this region is dominated by stochastic effects since $\phi_{\text{sr1}} < \phisto$. In the stochastic regime, the expressions we have previously derived receive a contribution from $\phi > \phi_{\text{sr1}}$ since we are integrating beyond $\phisto$, and are therefore inconsistent in this case. Intuitively, one may think that the violation of slow roll induces the suppression of the noise amplitude in the Langevin equation~(\ref{eq:intro:Langevin}) so that $\Delta\phiwell$ should in fact be replaced with $\phi_{\text{sr1}}$. The situation is then similar to the one sketched in \Fig{fig:sketch2}, and from \Eqs{eq:def:mu} and~(\ref{constraint3}), one has
\bea
\mu^2 \sim \frac{\phi_{\text{sr1}}^2}{v_0\Mp^2} = \frac{1}{v_0}\left(\frac{\phi_0}{\Mp}\right)^{\frac{2p}{p-2}}\, .
\eea
Since the condition under which this case is defined is $\phi_{0}/\Mp < v_{0}^{\frac{p - 2}{2p}}$, one has $\mu^2<1$, and PBHs are not overproduced in this case.
\subsection{Case $1 < p < 2$}
In this case, slow roll is valid in the range
\bea
\phi_{\text{sr1}} \ll \phi \ll \phi_{\text{sr2}}\, .
\eea
Note that this condition implies that $\phi_{\text{sr1}} \ll \phi_{\text{sr2}}$, which is the case only if $\phi_0\gg \Mp$. One then has $\phi_{\text{sr2}}\gg \phi_0$, so this range extends beyond the vacuum-dominated regime, and the interval of interest is in fact
\bea
\label{eq:app:example:case_1_lt_p_lt_2:sr_cond}
\phi_{\text{sr1}} \ll \phi \ll \phi_0\, .
\eea
Two sub-cases need to be distinguished.
\subsubsection{Case $1 < p < 2$ and $\phi_0/\Mp < v_{0}^{\frac{p - 2}{2p}}$}
In this case, \Eqs{constraint3}, \eqref{constraint1}, \eqref{app:example:Deltaphiwell} and \eqref{clpower>1} give rise to
\bea
\phicl < \phisto < \phi_{\text{sr1}} < \phi_0< \phi_{\text{sr2}}\, .
\eea
This means that in the region of interest given by \Eq{eq:app:example:case_1_lt_p_lt_2:sr_cond}, the classical approximation is valid and predicts that PBHs are not overproduced.
\subsubsection{Case $1 < p < 2$ and $\phi_0/\Mp > v_{0}^{\frac{p - 2}{2p}}$}
In this case, \Eqs{constraint3}, \eqref{constraint1}, \eqref{app:example:Deltaphiwell} and \eqref{clpower>1} give rise to
\bea
 \phi_{\text{sr1}} < \phisto < \phicl < \phi_0< \phi_{\text{sr2}}\, .
\eea
The situation is therefore similar to the case $p > 2$ and $\phi_{0} > \Mp$ described in \Sec{app:example:p_gt_2_and_phi0_gt_Mp}, and quantum diffusion does not sufficiently suppress PBH production.
\subsection{Case $0 < p < 1$}
If one takes $p<1$, contrary to the previous cases, the condition for the classical power spectrum to be larger than one is $\phi > \phicl$, where $\phicl$ is given by \Eq{clpower>1}. Furthermore, in this case, the slow-roll conditions read $\phi \gg \phi_{\mathrm{sr1}}$ and $\phi \gg \phi_{\mathrm{sr2}}$, and which of these conditions is the strictest depends on whether $\phi_0$ is sub- or super-Planckian.
\subsubsection{Case $0 < p < 1$ and $\phi_0<\Mp$}
In this case the slow-roll condition reads
\bea
\phi \gg \phi_{\mathrm{sr}1}\, .
\eea
Since $p<1$, $\phi_0<\Mp$ implies that $\phi_0/\Mp < v_0^{\frac{p-2}{2p}}$ and from \Eqs{constraint3}, \eqref{app:example:Deltaphiwell} and \eqref{clpower>1}, one has
\bea
\phisto < \phi_{\text{sr1}}  < \phicl   \, .
\eea
In this case, stochastic effects cannot play an important role in the slow-roll region of the potential, where the power spectrum is smaller than one provided $\phi< \phicl$.
\subsubsection{Case $0 < p < 1$ and $\phi_0>\Mp$}
In this case the slow-roll condition reads
\bea
\phi \gg \phi_{\mathrm{sr}2}\, ,
\eea
and two sub-cases need to be distinguished.
\paragraph{Case $0 < p < 1$ and $\frac{\phi_{0}}{\Mp} > v_{0}^{\frac{p - 2}{2p}}$}
$ $\\
From \Eqs{constraint1}, \eqref{app:example:Deltaphiwell} and \eqref{clpower>1}, one has
\bea
\phicl < \phi_{\text{sr2}} < \phisto \, .
\eea
In the classical region of the potential, $\phi>\phisto$, the power spectrum is much larger than one and PBHs are overproduced. In the stochastic region of the potential, $\phi_{\text{sr2}} <\phi< \phisto$, assuming $\phi_{\text{sr2}} \ll \phisto$, $\mu^2$ is given by \Eq{eq:app:example:mu}, which is much larger than one for $\phi_0>\Mp$ and $p<2$. Therefore PBHs are also too abundant in this part of the potential, and this case is observationally excluded.
\paragraph{Case $0 < p < 1$ and $1<\frac{\phi_{0}}{\Mp} < v_{0}^{\frac{p - 2}{2p}}$}
$ $\\
From \Eqs{constraint1}, \eqref{app:example:Deltaphiwell} and \eqref{clpower>1}, one has
\bea
\phisto < \phi_{\text{sr2}} < \phicl \, .
\eea
This case is similar to the one where $0<p<1$ and $\phi_0<\Mp$, and stochastic effects do not play an important role in the slow-roll region of the potential. 
\subsection{Case $p=1$}
Finally, let us consider the case $p=1$. The slow-roll approximation is valid throughout the entire vacuum-dominated region of the potential provided
\bea
\phi_0\gg\Mp.
\eea
This case is however more subtle than the previous ones, since from \Eq{eq:eta_class}, one has 
\bea
\eta_{\mathrm{class}}\equiv 0\, .
\eea
This means that the first stochastic correction in the classical expansion presented in \Sec{sec:ClassicalLimit} vanishes. This does not imply that quantum diffusion never plays a role since higher-order terms can still spoil the classical result, but this suggests that our classicality criterion fails in this case to identify where stochastic effects become important.

This is why no clear conclusion can be drawn in this case. In practice, one should extend the classical expansion of \Sec{sec:ClassicalLimit} to next-to-next-to-next to leading order (NNNLO) at least to determine what the first stochastic correction is and under which condition the classical approximation holds, and investigate numerically the cases where it does not. We leave these considerations for future work.

\bibliographystyle{JHEP}
\bibliography{PBH}
\end{document}

%% file: newcommands.tex
\newcommand{\ie}{{i.e.~}}

\newcommand{\eg}{e.g.~}

\newcommand{\phiend}{\phi_{\text{end}}}
\newcommand{\phiwell}{\phi_{\text{well}}}		
\newcommand{\phiuv}{\phi_{\text{uv}}}			

\newcommand{\phisto}{\Delta\phiwell}
\newcommand{\phicl}{\phi_{\mathcal{P}_{\zeta}|_{\mathrm{cl}}>1}}		
\newcommand{\N}{\mathcal{N}}

\let\oldsqrt\sqrt
\def\sqrt{\mathpalette\DHLhksqrt}
\def\DHLhksqrt#1#2{%
\setbox0=\hbox{$#1\oldsqrt{#2\,}$}\dimen0=\ht0
\advance\dimen0-0.2\ht0
\setbox2=\hbox{\vrule height\ht0 depth -\dimen0}%
{\box0\lower0.4pt\box2}}

\newcommand{\mean}[1]{\left\langle #1 \right\rangle}

\DeclareMathOperator{\erfc}{erfc}

\newcommand{\dd}{\mathrm{d}}
\newcommand{\ee}{e}

\newcommand{\sss}[1]{{\scriptscriptstyle{#1}}}

\newcommand{\uPl}{\mathrm{Pl}}
\newcommand{\uin}{\mathrm{in}}

\newcommand{\uend}{\mathrm{end}}

\newcommand{\uc}{\mathrm{c}}

\newcommand{\lo}{\sss{\mathrm{LO}}}
\newcommand{\nlo}{\sss{\mathrm{NLO}}}
\newcommand{\nnlo}{\sss{\mathrm{NNLO}}}

\newcommand{\usssPl}{\sss{\uPl}}

\newcommand{\uNL}{\mathrm{NL}}

\newcommand{\calP}{\mathcal{P}}

\newcommand{\Mp}{M_\usssPl}

\newcommand{\fnl}{f_\uNL}

\newcommand{\efolds}{$e$-folds}
\newcommand{\efold}{$e$-fold}

\newcommand{\beq}{\begin{equation}}
\newcommand{\eeq}{\end{equation}}
\newcommand{\bea}{\begin{equation}\begin{aligned}}
\newcommand{\eea}{\end{aligned}\end{equation}}

\newlength{\wsingfig}
\newlength{\wdblefig}
\newlength{\wquadfig}
\newlength{\wtriplefig}
\newcommand{\Eq}[1]{Eq.~(\ref{#1})}
\newcommand{\Eqs}[1]{Eqs.~(\ref{#1})}
\newcommand{\Fig}[1]{Fig.~{\ref{#1}}}
\newcommand{\Figs}[1]{Figs.~{\ref{#1}}}
\newcommand{\Ref}[1]{Ref.~{\cite{#1}}}
\newcommand{\Refs}[1]{Refs.~{\cite{#1}}}
\newcommand{\Sec}[1]{Sec.~\ref{#1}}
\newcommand{\Secs}[1]{Secs.~\ref{#1}}
\newcommand{\App}[1]{Appendix~\ref{#1}}